\documentclass[secnumarabic,amssymb, nobibnotes, aps, prl,preprint]{revtex4-1}
\usepackage{graphicx}	
\usepackage{graphics}	
\usepackage{amsmath}
\pdfoutput=1
\begin{document}
\title{Are Polaritons Strongly Interacting? Direct Measurement of the Polariton-Polariton Interaction Strength}
\author{Yongbao Sun,${}^1{}^\ast$ Yoseob Yoon,${}^1$ Mark Steger,${}^2$ Gangqiang Liu,${}^2$\\Loren N. Pfeiffer,${}^3$ Ken West,${}^3$ David W. Snoke,${}^2{}^\ast$  and Keith A. Nelson${}^1$}
\affiliation{${}^1$Department of Chemistry and Center for Excitonics, Massachusetts Institute of Technology, 77 Massachusetts Avenue, Cambridge, MA 02139, USA \\ 
${}^2$Department of Physics, University of Pittsburgh, 3941 O'Hara St., Pittsburgh, PA 15218, USA\\
${}^3$Department of Electrical Engineering, Princeton University, Princeton, NJ 08544, USA
}
\maketitle

\textbf{Exciton-polaritons in a microcavity are composite two-dimensional bosonic quasiparticles, arising from the strong coupling between confined light modes in a resonant planar optical cavity and excitonic transitions, typically using excitons in semiconductor quantum wells (QWs) placed at the antinodes of the same cavity.  Quantum phenomena such as Bose-Einstein condensation (BEC) \cite{Kasprzak2006,Balili2007}, superfluidity \cite{Amo2009}, quantized vortices \cite{Lagoudakis2008,Lagoudakis2009,Sanvitto2010,Nardin2011,Tosi2012}, and macroscopic quantum states \cite{Dreismann2014,Liu2015} have been realized at temperatures from tens of Kelvin up to room temperatures \cite{Christopoulos2007,Kena-Cohen2010,Plumhof2014}, and polaritonic devices such as spin switches \cite{Amo2010} and optical transistors \cite{Ballarini2013} have also been reported. Many of these effects of exciton-polaritons depend crucially on the polariton-polariton interaction strength. Despite the importance of this parameter, it has been difficult to make an accurate experimental measurement, mostly because of the difficulty of determining the absolute densities of polaritons and bare excitons.  Here we report the direct measurement of the polariton-polariton interaction strength in a very high-Q microcavity structure. By allowing polaritons to propagate over 40 $\mu$m to the center of a laser-generated annular trap, we are able to separate the polariton-polariton interactions from polariton-exciton interactions. The interaction strength is deduced from the energy renormalization of the polariton dispersion as the polariton density is increased, using the polariton condensation as a benchmark for the density. We find that the interaction strength is about two orders of magnitude larger than previous theoretical estimates, putting polaritons squarely into the strongly-interacting regime. When there is a condensate, we see a sharp transition to a different dependence of the renormalization on the density, which is evidence of many-body effects. }

Much of the physics of polaritons is dominated by the fact that they have extremely light effective mass. When an exciton is mixed with a cavity photon to become an exciton-polariton, it has an effective mass about four orders of magnitude less than a vacuum electron, and about three orders of magnitude less than a typical semiconductor quantum well exciton. (The Supplementary Information gives a basic introduction to the properties of exciton-polaritons.) Therefore one can view the polaritons as excitons which are given much longer diffusion length, with propagation distance of polaritons up to millimeters \cite{Steger2013,Steger2015}; this may have implications for solar cells, which depend crucially on the diffusive migration of excitons \cite{Zhugayevych2015}. Alternatively, exciton-polaritons can be viewed as photons with nonlinear interactions many orders of magnitude higher than in typical optical materials, due to their excitonic components \cite{Snoke2012}. The light effective mass of the polaritons (typically $10^{-8}$ that of a hydrogen atom) allows for quantum phenomena to be realized at much higher temperatures than in cold atomic gases. \\
\\
\noindent\textbf{Interactions among polaritons.}  The interaction of exciton-polaritons is presumed to come exclusively from their underlying excitonic components. These interactions renormalize the polariton dispersion, leading to a continuous blue shift to higher emission energies as the polariton density increases. To lowest order, one can view the blue shift of the polariton emission line as the real self-energy due to the interactions, and the spectral width of the polariton emission line as the imaginary self-energy arising from the same interactions. When the particles undergo Bose-Einstein condensation, the spectral line width narrows due to the spontaneously emerged coherence. 

Most previous theoretical work has assumed the exciton-exciton interaction strength calculated by Tassone and Yamamoto \cite{Tassone1999}, found to be $g\sim6E_Ba_B^2$, where $E_B$ is the excitonic binding energy, and $a_B$ is the excitonic Bohr radius. Using the typical values for excitons in GaAs narrow quantum wells, namely $E_B\sim10$ meV and $a_B\sim100 $ \AA, gives $g\sim 6$ $\mu$eV$\cdot \mu$m$^2$. This, in turn, implies that the polariton gas is intrinsically weakly interacting, as measured by the unitless parameter $\gamma = g/(\hbar^2/2m)\sim 0.01$ \cite{Prokofev2001,Prokofev2002}, using the effective mass of polaritons $m\sim10^{-4}m_0$, where $m_0$ is the vacuum electron mass.

 \begin{figure}[htbp]
 \centering
   \includegraphics[width=1\textwidth]{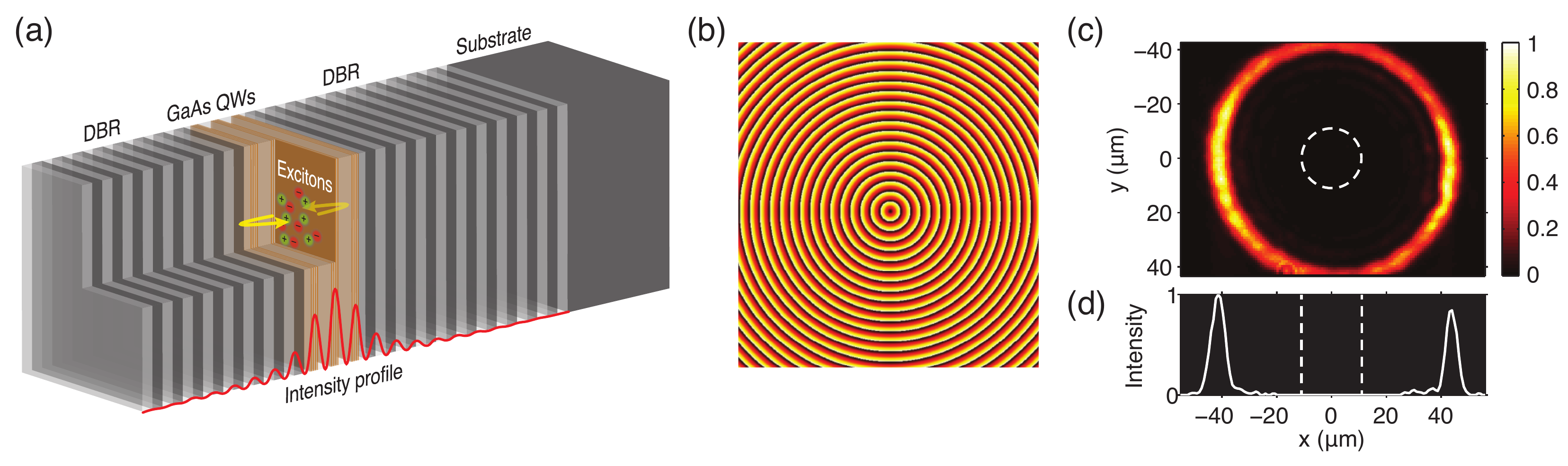}
    \caption{ (color online). (a) Long-lifetime microcavity structure used in the experiment. The dark and light gray-colored layers are the distributed Bragg reflectors (DBR) used to confine the light.  The red solid line indicates the intensity distribution of the confined cavity mode. The brown-colored layers indicate the position of the quantum wells (QWs), with green and red dots showing the bounded electron-hole pair, i.e., excitons. The yellow arrows indicate the energy oscillation between the cavity mode and the exciton states. (b) An representative axicon phase imprinted on the surface of the spatial light modulator (SLM) used to generate the annular excitation pattern. (c) The real-space image of the laser reflection, which is a ring with a diameter of 85 $\mu m$, and the white dashed circle indicates the region where the photoluminescence (PL) is collected. (d) White solid line is the cross section of $y = 0$ slice that is selected by the entrance of the imaging spectrometer and then energy-resolved. The white dashed line indicates the position of the pinhole. }
     \end{figure}

Under the mean-field approximation, at low particle density, the blue shift of ground state energy is linearly dependent on the  particle density $n$, i.e.,
\begin{equation}
\Delta E = gn \label{mft}
\end{equation}
where $\Delta E$ and $n$ are the blue shift and particle density, respectively. In the absence of many-body correlations, the slope $g$ is density independent. In the experiments reported here, we measured the energy shift $\Delta E$ of the polariton ground state, which has in-plane momentum $k_{||}=0$, for various excitation densities. The interaction strength was then extracted as the slope of the energy shift with density.

The main difficulty in previous experiments for measuring the polariton-polariton interacting strength has been to establish the absolute density of the polaritons independent of any excitons that may be present. When the polaritons are generated by non-resonant laser excitation, not only polaritons but also bare excitons are created. This leads to a population of excitons of unknown density. If structures with short lifetime (ca.~2 ps) are used, the polaritons do not travel far from the cloud of excitons. Therefore, when a shift of the energy of the polaritons is seen, it is not possible to determine how much of this shift is due to direct polariton-polariton interaction and how much is due to polariton-exciton interaction. In principle, there may also be free carriers created near the laser excitation region, which also interact with the polaritons. Alternatively, if polaritons are created resonantly, they typically have a highly non-equilibrium and coherent distribution that does not allow easy extraction of the energy renormalization of the particles. 

In this work, we used a very high-$Q$ microcavity under non-resonant excitation. The high $Q$ implies long polariton lifetime, 200 ps or longer \cite{Steger2015}, which allows the polaritons to propagate well away from the laser excitation region \cite{Steger2013}. By propagating polaritons to the center of an sufficiently big optically-induced annular trap, we can separate the polaritons from the population of free carriers and hot excitons. We can then extract the polariton interaction strength from the spectral data at the center of the trap. 

The interactions between polaritons are predicted to be spin-dependent, and several experiments have studied this spin dependence \cite{Amo2010,Anton2015,Sich2014,Sekretenko2013,Takemura2014_2}. In the experiment reported here, we assume that the polaritons are equilibrated into a mixture of both spin states \cite{Kasprzak2006}, and therefore the interaction strength we measure is an average value. Because the interaction of polaritons is stronger for spin-aligned polaritons, we expect that our measured interaction strength will be dominated by the spin-aligned contribution. \\
\\
\noindent\textbf{Experimental methodologies}.
The sample design is an Al$_{0.2}$Ga${}_{0.8}$Al/AlAs $3\lambda/2$ optical cavity, with four groups of 7 nm GaAs QWs embedded at each of the three antinodes, as shown in Fig.~1a. It has a quality factor of $\sim$$320,000$ and a cavity lifetime of $\sim$$135$ ps, which is significantly higher than most samples used in earlier experiments (the lifetimes of polaritons were at most 30 ps \cite{Wertz2010}, even well into excitonic detunings), confirmed by the propagation of resonantly created polaritons up to millimeter scale \cite{Steger2015}. The polariton lifetime is proportional to its excitonic fraction, which in turn depends on the cavity detuning $\delta$, defined as the energy difference between the bare photon energy at $k_{||}=0$ and the bare exciton energy. Because the microcavity structure used in this work is wedged, with a photon energy gradient of $\sim$$11$ meV/mm, the detuning is then a function of the location on the microcavity structure. We can easily vary the detuning, and therefore the excitonic and photonic fractions of the polaritons, by doing the experiment at different sample positions on the microcavity. Careful calibration of the upper and lower polariton energies as a function of position on the sample allows us to determine the detuning, and therefore the excitonic and  photonic fractions of the polaritons, at each position. At the resonance of the excitons and cavity photons, the lifetime of polaritons in our sample is $\sim$270 ps. 

During the experiment, the sample was thermally attached to a cold finger in an open-loop cryostat which was stabilized at 10 K. In order to accumulate polariton densities after they accelerate away from the excitation region, we made a spatial trap.  We made use of the fact that excitons created by a laser field in the same structure have much larger effective mass than the polaritons; the typical polariton mass is less than $10^{-4}$ times the electron mass, while the free exciton mass is of the order a tenth of the electron mass in GaAs quantum wells. Therefore, the excitons created by a laser pulse are essentially static as seen by polaritons, and the mean-field repulsion energy of the excitons acts as a static potential energy barrier. This method has been used previously in several experiments to confine polaritons \cite{Tosi2012_2,Cristofolini2013, Askitopoulos2013, Askitopoulos2015}.

The annular trap in this work was formed by irradiating a spatial light modulator (SLM) with a pixel density of 1920$\times$1080 by a MSquared continuous wave (c.w.) laser. The ability to address each individual pixel on the SLM allows for shaping the incident Gaussian beam into any designed intensity patterns \cite{Cristofolini2013, Askitopoulos2013, Askitopoulos2015}. We imposed an axicon phase map, shown in Fig.~1b, onto the surface of the SLM which was placed at the Fourier plane of an imaging lens. At the conjugate plane of this lens, the Gaussian beam is transformed into a ring-shaped pattern with a diameter of $\sim$2 mm. With another two pairs of telescopes, a ring with a diameter of 85 $\mu m$ was formed on the sample surface, as shown in Fig~1c. The wavelength of the c.w. laser was tuned to 720 nm, about 130 meV above the lower polariton resonance, to match the second reflection minimum above the stop band of the cavity at resonance. This generated a high density of free carriers along the ring. In order to avoid any unwanted sample heating, the c.w. laser was modulated by an acousto-optic modulator (AOM) at 1 kHz with a duty cycle of 0.5\%. 
\begin{figure}[htbp]
 \centering
   \includegraphics[width=0.85\textwidth]{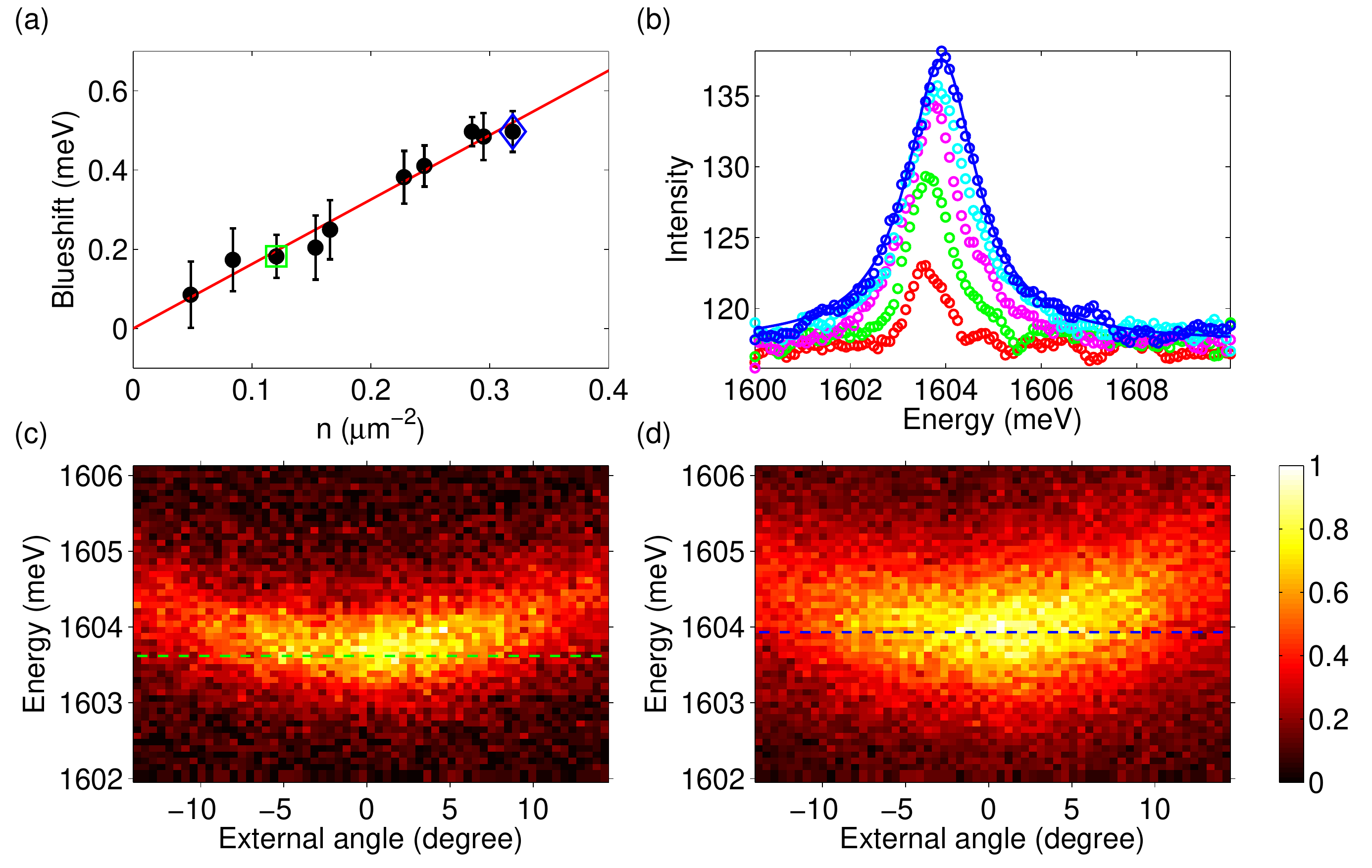}
    \caption{(color online) (a) Blue shift of ground state energies at different polariton densities at a detuning of $\Delta = 10.11$ meV when the excitation annular profile has a diameter of 85 $\mu$m. (b) Polariton emission profiles at $k_{||}=0$ at different densities. The blue and red lines correspond to the densities shown as blue diamond and green square in (a). The top emission profile was fit to a Lorentzian linewidth profile with a width of $\gamma = 0.89\pm0.03$ meV, as shown by the solid blue line. (c)-(d) Typical energy-resolved dispersions corresponding to the data points indicated by the green square (c) and blue diamond (d) in (a), both at low density when there is no condensate. The dashed lines are the assigned energies of the ground state of polaritons. Colors are shown to calibrate the relative magnitude of the emission intensities.}
 \end{figure}

The photoluminescence (PL) from the non-resonant excitation was collected by a microscope objective with a numerical aperture (N.A.) of 0.28, and was imaged in the far-field geometry to the entrance slit of a spectrometer CCD. A spatial filter was placed at a reconstructed real-space plane downstream to filter out the emission $11$ $\mu $m away from the center of the trap. The dashed circle in Fig.~1c and the dashed line in Fig.~1d indicates the collection region. The $y = 0$ slice was selected by the spectrometer slit and then spectrally dispersed to get the emission energy. Fig.~1d shows the intensity profile of the $y=0$ slice. We also measured the integrated light intensity in the collection region with and without a neutral density filter, and determined the light intensity in the collection region is $\sim$2000 times smaller than that of the pumping region. This is crucial to the quality of the measurement, for otherwise, light in the collection region can create excitons and free carriers which can lead to an additional shift of the polariton energy. As discussed below, typical energy shifts of the polaritons in the collection region are about 10\% of the energy shift in the pumping region.  Assuming that the exciton density is proportional to the pump intensity, the energy shift of the polaritons in the collection region cannot come from the excitons. 

Fig.~2c shows a typical energy-resolved emission pattern in far-field geometry (i.e., angle-resolved PL, which gives the momentum distribution of the polaritons) at low excitation power, and Fig.~2d shows a case of high excitation power, but below the condensation threshold. As can be seen, the ground state of polariton energies, indicated by the horizontal dotted and dashed lines, are blue shifted due to repulsive interactions among polaritons. As seen in Fig.~2b, there is also a Lorentzian spectral broadening, which also arises from the interactions. To a good approximation, when there is no condensate, this spectral broadening can be seen as simply the uncertainty energy given by $\Delta E=\hbar/\Delta t$, where $\Delta t$ is the average incoherent scattering time.  As discussed below, this broadening also gives a measure of the polariton-polariton interaction strength. At low density, the emission profile deviates away from the Lorentzian profile and resembles the Voigt profile, which accounts for the disorder broadening, as discussed in the Supplementary Information.\\
\\
\noindent\textbf{Calibration of polariton density}.
Care was taken to ensure that the polaritons were nearly in spatial and thermal equilibrium. The Supplementary Information presents images of the spatial distributions of the polaritons in a trap under similar conditions, showing that the distributions are nearly spatially homogeneous. At high density when a condensate forms, there is evidence of self-trapping into a central region of the laser-generated trap. The data reported here are for detunings with significant exciton component, allowing good thermalization; the Supplementary Information shows typical energy distributions of the polaritons at low densities. When polaritons are very photon-like, the spatial and energy profiles of the polaritons become quite inhomogeneous; this will be the subject of another publication \cite{phase_boundaries}.

The greatest uncertainty in these measurements is the determination of the polariton density, i.e., the total number of polaritons in the area of the field of view. We used two methods and found consistency between them. The first was to carefully determine the absolute collection efficiency of our photon detection system, as discussed in the Supplementary Information. The number of polaritons at a given momentum $k_{||}$ was then deduced from the rate of photon emission knowing the cavity photon lifetime and the photonic fraction of the polaritons at that $k_{||}$.

The second method allowed us to set an absolute density by pegging the density at which the distribution of the polaritons is altered by the Bose-Einstein statistics of the particles. As discussed above, the polaritons are nearly in thermal equilibrium at resonance up to excitonic detunings. The equilibrium Bose-Einstein distribution is given by
\begin{equation}
N(E) = \frac{1}{e^{(E-\mu)/k_BT}-1}\label{bec}
\end{equation}
At densities well below the condensation threshold, this becomes a Maxwell-Boltzmann distribution $N(E)\propto e^{-E/k_BT}$. In the quantum regime, however, when $N(E)\sim1$ for low-energy states, the shape of the distribution is changed, and can be fit to a value of the chemical potential $\mu$. The value of $\mu$ controls not only the shape, but also the absolute value of $N(E)$. Therefore we have a tight constraint on the values of $\mu$ when we fit $N(E)$ at several particle densities. By minimizing the mean-squared error in the fitting of a set of $n$ distributions $N(E)$ collected at different pumping powers to the equilibrium Bose-Einstein model as in Eq. (\ref{bec}) with $2n+1$ free parameters, we deduce one single collection efficiency $\xi$ as well as $n$ temperatures and $n$ chemical potentials for different distributions. Detailed data analysis routine is included in the Supplementary Information. This collection efficiency was within 10\% of the number deduced by estimating the absolute photon collection efficiency, discussed above. The Supplementary Information includes representative thermalized distributions and the corresponding Bose-Einstein fits.

These methods give us an accurate calibration of the polariton density in the field of view in the annular trap. More generally, we do not have to rely on the details of the above calculations to get the approximate range of the polariton densities.  We know that we measure the polariton spectra up to the condensation threshold (as evidenced by the sharp spectral narrowing, discussed below and also reported in Ref. \cite{Nelsen2013}). Quantum effects will be important when the thermal de Broglie wavelength is comparable to the interparticle spacing, i.e., when
\begin{equation}
\lambda_T = \sqrt{\frac{2\pi\hbar^2}{mk_BT}}\sim r_s = n^{-1/2}
\end{equation}
where $n$ is the particle density, and $T$ is temperature of the particle. This implies $n\sim mk_BT/2\pi\hbar^2\approx4\times10^7$ cm$^{-2}$ near the condensation threshold, for $k_BT\sim 20$  K and the polariton mass given above.\\
\\
\noindent\textbf{Interaction strength of polaritons at low particle densities below the condensation threshold}. The measured number of polaritons increased linearly with pump excitation density in the low-density regime below the condensation threshold. In this range of densities, the ground state energies of the polaritons were extracted as shown in Fig.~2c and 2d. In order to determine the ground state energy at zero excitation density, we linearly extrapolate the measured blue shifts to zero density. All the blue shifts are then reported with respect to this energy. As shown in Fig.~2a, the blue shift increases linearly with the polariton density. This confirms that the interaction strength $g$ does not depend on the density of the polaritons in this regime.

\begin{figure}[htbp]
\begin{center}
  \includegraphics[width=0.5\textwidth]{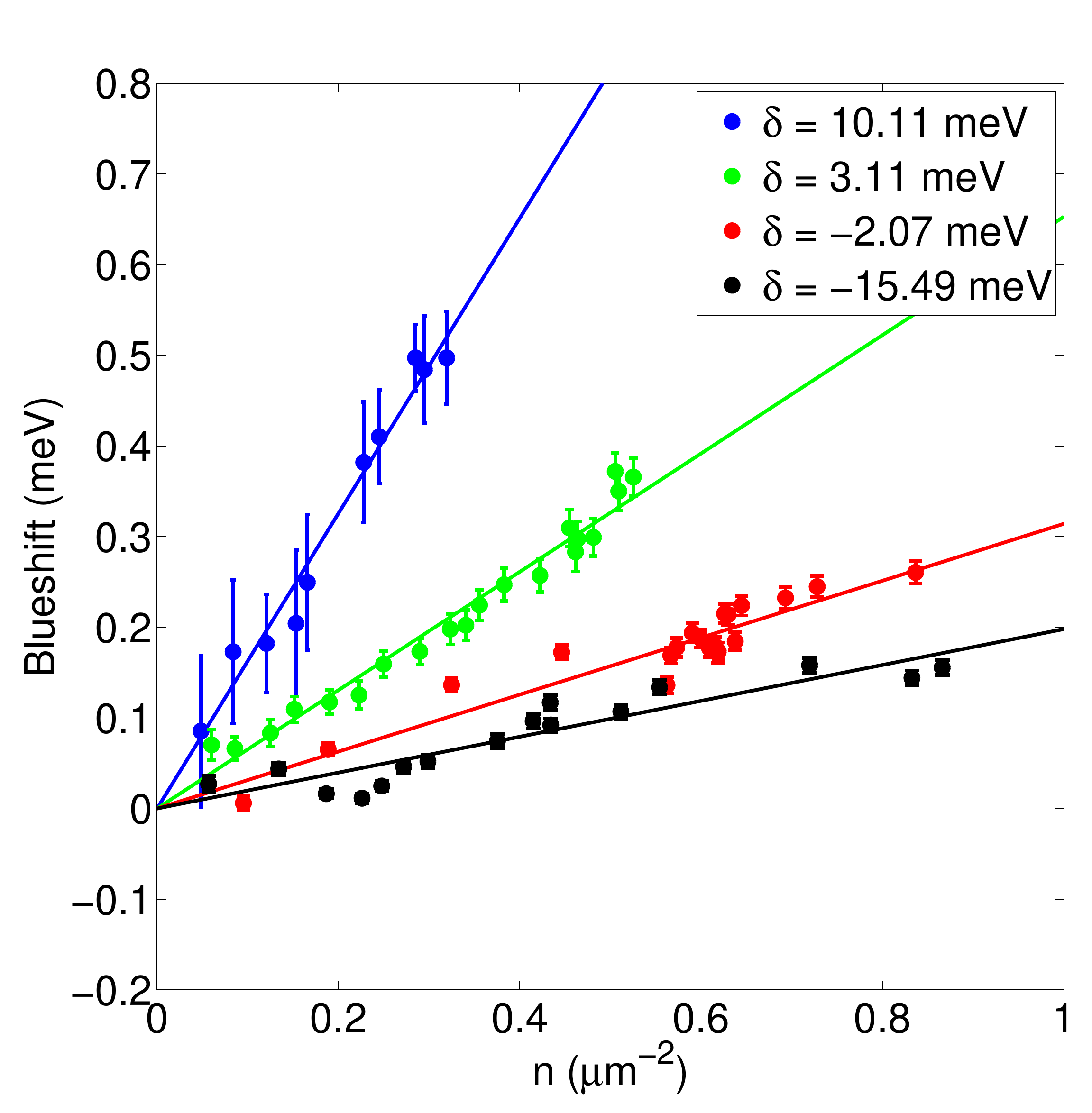}
   \caption{(color online) Blue shifts of the ground state energies as a function of polariton density at different cavity detunings. The solid points are measured blue shifts, the linear lines are fits to $\Delta E= gn$ using weighted least-squares estimates.}
\end{center}
\end{figure}

In Fig.~3, we plot the measured blue shifts of the $k_{||}=0$ state as a function of polariton density at four representative detunings. As expected, polaritons experience higher blue shifts for excitonic detunings than for photonic detunings, since the energy renormalization presumably comes from the excitonic constituents. The solid lines show linear fits to the data using weighted least square estimates, as discussed in the Supplementary Information.

Fig.~4 shows the extracted slopes at different cavity detunings versus their corresponding excitonic fractions. The interaction strength increases when excitonic fractions in polaritons are higher. The standard theory predicts that the interaction between polaritons is governed by their underlying excitonic fractions $x$, given by
\begin{equation}
x = \frac{1}{2}\left(1+\frac{\delta}{\sqrt{\delta^2+\Omega^2}}\right),\label{hf}
\end{equation}
where $\Omega$ is the full Rabi splitting and $\delta$ is the cavity detuning, defined as $\delta = E_c(k_{||}=0)-E_x$, where $E_c$ and $E_x$ are the cavity and exciton energies, respectively. The solid line is a quadratic fit to the extracted slopes in Fig.~3. The dependence of the shift on exciton fraction is clearly superlinear, which is another indicator that the shift is not arising from polariton-exciton interactions; if interaction with excitons were the dominant cause of the blue shift, the shift would be linear with exciton fraction. 

By extrapolation, the interaction strength in the limit of $x \rightarrow 1$ is determined as $1.74\pm0.46$ meV$\cdot\mu$m$^2$.
 \begin{figure}[htbp]
 \centering
   \includegraphics[width=0.5\textwidth]{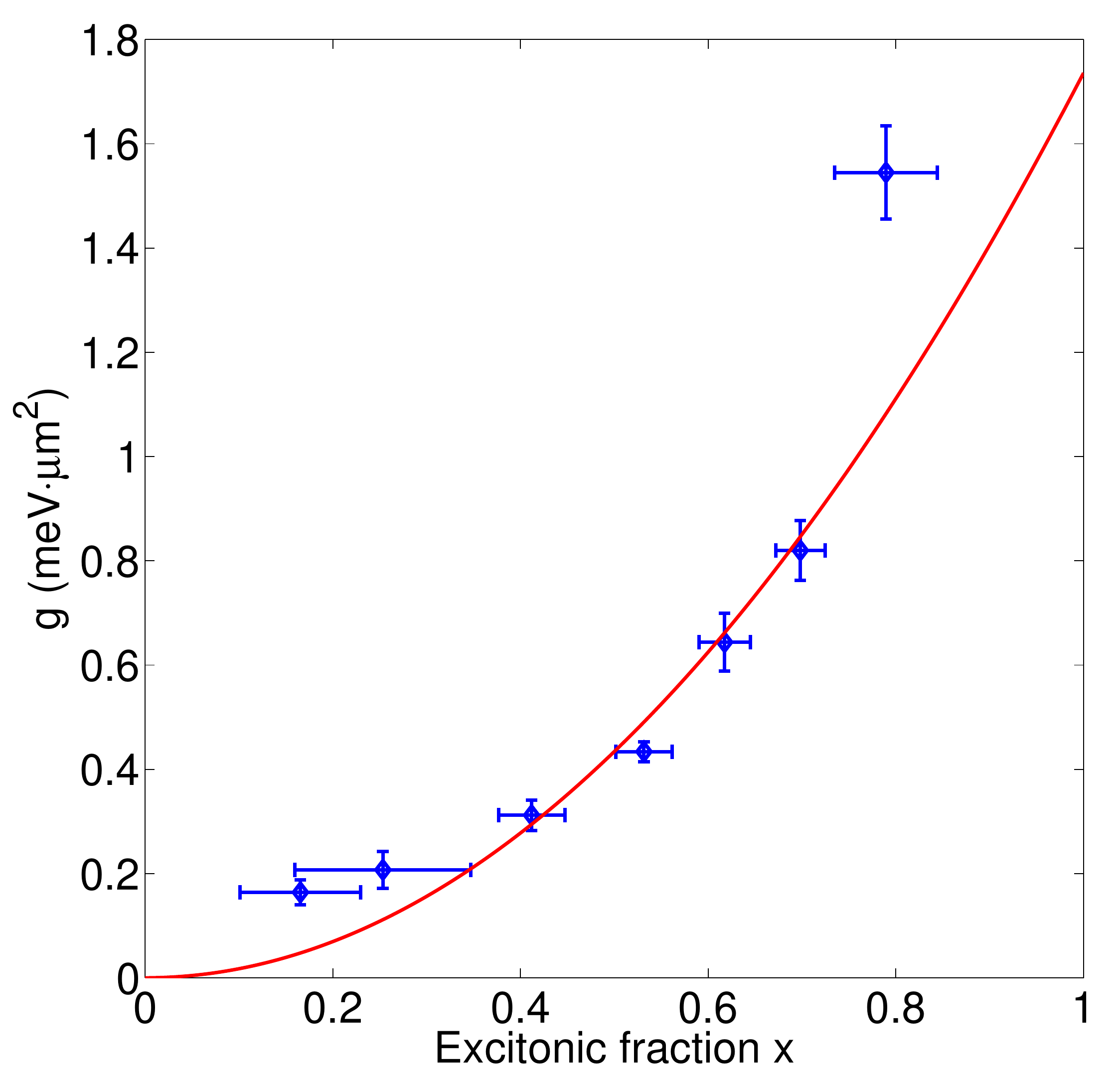}
    \caption{(color online) Measured interaction strength ($g$) as a function of excitonic fraction ($x$), which is deduced for each data set from the effective mass of the lower polariton dispersion using the formula $x \simeq 1- m_c/m_{LP}$, where $m_c$ is the bare cavity photon mass, measured at very large photonic detuning ($-$22.5 meV). Solid points are slopes extracted from Fig.~3. The red line is a best fit to a quadratic dependence using weighted least-squares estimates. }
 \end{figure}
This value is quite surprising, in that it is two orders of magnitude larger than the absolute value calculated in Ref.~\cite{Tassone1999}, which has guided many theoretical investigations of polaritons. It implies that the unitless parameter for the strength of the interactions is $\gamma=g/(\hbar^2/2m)\sim1$, which means that the polariton gas cannot be treated by standard weakly interacting Bose gas theory. In particular, it implies that many-body correlations will play an important role at high densities.

The extrapolation to $x \rightarrow 1$ seems to imply that not only polaritons, but bare Wannier excitons also have this strong interaction. However, we note that in all the measurements here, the effective mass of the particles was very light, comparable to the polariton mass at resonance (see the Supplementary Information for a plot of the mass as a function of detuning). It may be that the strong interactions seen here depend crucially on the light effective mass of the polaritons, and therefore it may be unwarranted to extrapolate this result to the pure exciton limit. 

As noted above, the polariton-polariton interactions affect not only the blue shift (real self-energy) but also the line broadening (imaginary self-energy). The Lorentzian line broadening value of $\sim 1$~meV also is two orders of magnitude larger than that would be expected for the nominal exciton-exciton interaction strength. Using the semiclassical formula $\tau^{-1} = n\sigma \bar{v}$, where $n$ is the density, $\sigma \sim a_B$ is the scattering cross section, and $\bar{v}$ is the average thermal velocity. It gives $\tau \sim 100$~ps, or $\hbar/\tau \sim 0.01$~meV with typical values for microcavity polaritons $n=4\times10^{7}$ cm$^{-2}$, $\sigma=100$ \AA $ $ and $ $  $\bar{v}=2\times10^8$ cm/s for $T=20$ K. The measured Lorentzian broadening corresponds to a polariton-polariton scattering time of less than 1 ps. \\
\\
\noindent\textbf{Saturation of the interaction strength at high particle densities above the condensation threshold}.
Evidence of the importance of many-body correlations is seen in the behavior of the blue shift at the condensation threshold. In Fig.~5, we plot the blue shift of the ground state emissions at different polariton densities up to and above the condensation threshold in an annulus trap with a diameter of 72 $\mu$m, for an excitonic detuning of $\delta =4.31 $ meV. We also show the spectral width of the emission at $k_{||}=0$. As discussed above, the condensation threshold is indicated by spectral narrowing. The corresponding energy distributions $N(E)$ can be found in Fig.~S3 in the Supplementary Information.

At low densities, the blue shift scales linearly as the polariton density with a slope of $0.91\pm0.13$ meV$\cdot\mu$m$^2$, consistent with the value expected for this detuning. The linewidth of the emission also increases linearly with density because the larger population leads to more frequent incoherent collisions, and then narrows due to the emergence of the coherence when condensation occurs.  At the condensation threshold, the blue shift of the polariton emission becomes strongly sublinear with density, following a power law of $\Delta E \sim n^{0.037}$. The simple mean-field prediction for a weakly interacting Bose gas is that the shift should still be linear with density when it condenses, with an overall slope that is reduced by a factor of two due to the nature of quantum indistinguishability. The fact that the shift is strongly sublinear points to the importance of high-order correlations in the condensate regime, which have been shown to have a significant role in strongly interacting gases \cite{Zimmermann2006}. \\
\begin{figure}[htbp]
\centering
   \includegraphics[width=0.75\textwidth]{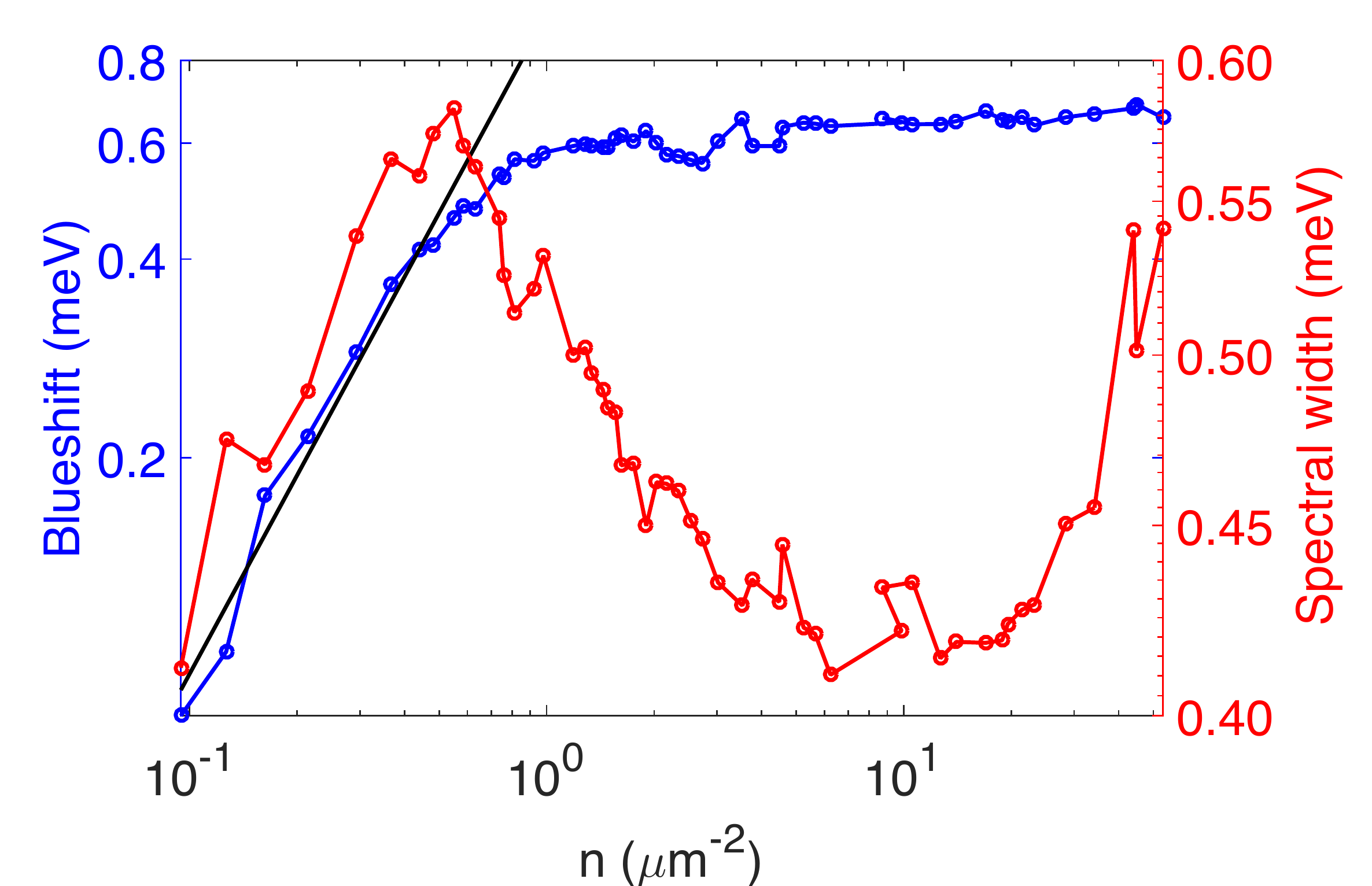}
    \caption{(color online) Measured blue shifts (blue circles) and spectral width (red circles) of the $k_{||}=0$ lower polariton emission as a function of polariton density at a cavity detuning of $\delta = 4.31$ meV in an annulus trap with a diameter of 72 $\mu$m. The straight line shows a linear dependence.}
 \end{figure}
\\
\noindent\textbf{Implications and outlook}. Given that the measured value of the blue shift is so much larger than the theoretical expectation, we consider possible ways that the measurement could be misleading. First, one may ask whether excitons and/or free carriers generated by the non resonant laser may indeed be diffusing into the region where the polaritons are observed. We are confident that this is not the case. As the diameter of the ring trap is increased up to 80~$\mu$m, the blue shift at fixed density has a constant value independent of the trap radius. (See the Supplemental Information for a plot of these data.) If exciton or free carrier diffusion were playing an important role, the blue shift at the condensate density would depend strongly on the ring radius, as their diffusion length becomes much less than the radius. Also, the exciton diffusion length has been directly measured (see, e.g., Ref.~\cite{Liu2015}, Figure 3a) and is less than 10~$\mu$m. The Supplementary Information shows data for the diffusion length of the excitons. Furthermore, it is also known that the diffusion length of the polaritons decreases with increasing excitonic component, which will lead to less exciton population far from the excitation region, while here we see that the blue shift increases with increasing excitonic fraction.

A blue shift may arise from phase-space filling, which leads to a reduction of the Rabi splitting between the upper and lower polariton branches. This is also predicted to give shifts of order $E_Ba_B^2$ or less \cite{saturation,pavlovic}, much less than what we observe here. Band-gap renormalization due to heating (i.e., phonon population) is known to give a red shift, not a blue shift, with increasing temperature.

Another possibility is that the polaritons generate free excitons in the center of the trap via thermal up-conversion. In this model, excitons are present in the center of the trap, not due to diffusion from the laser excitation region, i.e., from the walls of the ring trap, but from migration of polaritons by themselves, which then turn back into free excitons by thermal excitation. For this to give the blue shift we see, assuming the standard interaction strength, there would need to be two orders of magnitude more excitons than polaritons. This would severely deplete the population of polaritons.  Also, this effect should be exponentially dependent on the splitting $\Delta E$ between the polariton and exciton states, due to the Boltzmann $e^{\Delta E/k_BT}$ factor, but in fact we see even more blue shift than expected when the lower polaritons are very photonic, with large $\Delta E$.  We also see very little temperature dependence of the blue shift at constant $\Delta E$ as the lattice temperature is changed.

Assuming that the above mechanisms are ruled out, we can ask what mechanism could give us much stronger effective interactions at low density than predicted. One possibility is that the nearby biexciton resonance \cite{Takemura2014} significantly affects the interactions. We believe this is not the case, because the dependence of the interaction strength on detuning is not especially strong. Another possibility is a mechanism involving disorder, which has also been invoked as a way of explaining the larger-than-expected blue shift in previous experimental results \cite{Marchetti2006, Marchetti2007}. In our experiments, the photonic disorder is very low, but the excitonic disorder is approximately 2 meV (as measured by the inhomogeneous broadening of the bare exciton line). Jacob Taylor \cite{Taylor2015} has suggested that disorder may cause the excitons participating in the polariton states to be spatially correlated in ``puddles'' of much higher density than the average polariton density. This view requires no change to the standard exciton-exciton interaction strength. It is supported by the observation that both the blue shift and the Lorentzian line broadening are about 100 times larger than expected; both of these depend linearly on the effective density.

In conclusion, we have directly measured the polariton-polariton interaction strength by propagating polaritons to the center of a laser-generated annular trap. The interaction strength is independent of polariton density at low-density regime. The measured value is two orders of magnitude larger than expected from prior theoretical calculations. 
The polariton-polariton interaction is effectively a $\chi^{(3)}$ nonlinearity \cite{Snoke2012}, and is an important parameter for the design and theoretical modeling exciton-polaritonic optical devices. The large value measured here indicates that the efficiency of these devices may be much greater than anticipated, at least in the low-density regime.  The saturation of the  blue shift upon condensation calls for further theoretical analysis, and is likely due to coherent many-body effects of the condensate. 
\bibliographystyle{naturemag}
\bibliography{interactions_NP}

\begin{thebibliography}{10}
\expandafter\ifx\csname url\endcsname\relax
  \def\url#1{\texttt{#1}}\fi
\expandafter\ifx\csname urlprefix\endcsname\relax\def\urlprefix{URL }\fi
\providecommand{\bibinfo}[2]{#2}
\providecommand{\eprint}[2][]{\url{#2}}

\bibitem{Kasprzak2006}
\bibinfo{author}{Kasprzak, J.} \emph{et~al.}
\newblock \bibinfo{title}{Bose-einstein condensation of exciton polaritons}.
\newblock \emph{\bibinfo{journal}{Nature}} \textbf{\bibinfo{volume}{443}},
  \bibinfo{pages}{409--414} (\bibinfo{year}{2006}).

\bibitem{Balili2007}
\bibinfo{author}{Balili, R.}, \bibinfo{author}{Hartwell, V.},
  \bibinfo{author}{Snoke, D.}, \bibinfo{author}{Pfeiffer, L.} \&
  \bibinfo{author}{West, K.}
\newblock \bibinfo{title}{Bose-einstein condensation of microcavity polaritons
  in a trap}.
\newblock \emph{\bibinfo{journal}{Science}} \textbf{\bibinfo{volume}{316}},
  \bibinfo{pages}{1007--1010} (\bibinfo{year}{2007}).

\bibitem{Amo2009}
\bibinfo{author}{Amo, A.} \emph{et~al.}
\newblock \bibinfo{title}{Superfluidity of polaritons in semiconductor
  microcavities}.
\newblock \emph{\bibinfo{journal}{Nat. Phys.}} \textbf{\bibinfo{volume}{5}},
  \bibinfo{pages}{805--810} (\bibinfo{year}{2009}).

\bibitem{Lagoudakis2008}
\bibinfo{author}{Lagoudakis, K.~G.} \emph{et~al.}
\newblock \bibinfo{title}{Quantized vortices in an exciton-polariton
  condensate}.
\newblock \emph{\bibinfo{journal}{Nat. Phys.}} \textbf{\bibinfo{volume}{4}},
  \bibinfo{pages}{706--710} (\bibinfo{year}{2008}).

\bibitem{Lagoudakis2009}
\bibinfo{author}{Lagoudakis, K.~G.} \emph{et~al.}
\newblock \bibinfo{title}{Observation of half-quantum vortices in an
  exciton-polariton condensate}.
\newblock \emph{\bibinfo{journal}{Science}} \textbf{\bibinfo{volume}{326}},
  \bibinfo{pages}{974--976} (\bibinfo{year}{2009}).

\bibitem{Sanvitto2010}
\bibinfo{author}{Sanvitto, D.} \emph{et~al.}
\newblock \bibinfo{title}{Persistent currents and quantized vortices in a
  polariton superfluid}.
\newblock \emph{\bibinfo{journal}{Nat. Phys.}} \textbf{\bibinfo{volume}{6}},
  \bibinfo{pages}{527--533} (\bibinfo{year}{2010}).

\bibitem{Nardin2011}
\bibinfo{author}{Nardin, G.} \emph{et~al.}
\newblock \bibinfo{title}{Hydrodynamic nucleation of quantized vortex pairs in
  a polariton quantum fluid}.
\newblock \emph{\bibinfo{journal}{Nat. Phys.}} \textbf{\bibinfo{volume}{7}},
  \bibinfo{pages}{635--641} (\bibinfo{year}{2011}).

\bibitem{Tosi2012}
\bibinfo{author}{Tosi, G.} \emph{et~al.}
\newblock \bibinfo{title}{Geometrically locked vortex lattices in semiconductor
  quantum fluids}.
\newblock \emph{\bibinfo{journal}{Nat. Comm.}} \textbf{\bibinfo{volume}{3}},
  \bibinfo{pages}{1243} (\bibinfo{year}{2012}).

\bibitem{Dreismann2014}
\bibinfo{author}{Dreismann, A.} \emph{et~al.}
\newblock \bibinfo{title}{Coupled counterrotating polariton condensates in
  optically defined annular potentials}.
\newblock \emph{\bibinfo{journal}{Proc. Nat. Acad. Sci.}}
  \textbf{\bibinfo{volume}{111}}, \bibinfo{pages}{8770--8775}
  (\bibinfo{year}{2014}).

\bibitem{Liu2015}
\bibinfo{author}{Liu, G.}, \bibinfo{author}{Snoke, D.~W.},
  \bibinfo{author}{Daley, A.}, \bibinfo{author}{Pfeiffer, L.~N.} \&
  \bibinfo{author}{West, K.}
\newblock \bibinfo{title}{A new type of half-quantum circulation in a
  macroscopic polariton spinor ring condensate}.
\newblock \emph{\bibinfo{journal}{Proc. Nat. Acad. Sci.}}
  \textbf{\bibinfo{volume}{112}}, \bibinfo{pages}{2676--2681}
  (\bibinfo{year}{2015}).

\bibitem{Christopoulos2007}
\bibinfo{author}{Christopoulos, S.} \emph{et~al.}
\newblock \bibinfo{title}{Room-temperature polariton lasing in semiconductor
  microcavities}.
\newblock \emph{\bibinfo{journal}{Phys. Rev. Lett.}}
  \textbf{\bibinfo{volume}{98}}, \bibinfo{pages}{126405}
  (\bibinfo{year}{2007}).

\bibitem{Kena-Cohen2010}
\bibinfo{author}{Kena-Cohen, S.} \& \bibinfo{author}{Forrest, S.~R.}
\newblock \bibinfo{title}{Room-temperature polariton lasing in an organic
  single-crystal microcavity}.
\newblock \emph{\bibinfo{journal}{Nat. Photon.}} \textbf{\bibinfo{volume}{4}},
  \bibinfo{pages}{371--375} (\bibinfo{year}{2010}).

\bibitem{Plumhof2014}
\bibinfo{author}{Plumhof, J.~D.}, \bibinfo{author}{Stoferle, T.},
  \bibinfo{author}{Mai, L.}, \bibinfo{author}{Scherf, U.} \&
  \bibinfo{author}{Mahrt, R.~F.}
\newblock \bibinfo{title}{Room-temperature bose-einstein condensation of cavity
  exciton-polaritons in a polymer}.
\newblock \emph{\bibinfo{journal}{Nat. Mater.}} \textbf{\bibinfo{volume}{13}},
  \bibinfo{pages}{247--252} (\bibinfo{year}{2014}).

\bibitem{Amo2010}
\bibinfo{author}{Amo, A.} \emph{et~al.}
\newblock \bibinfo{title}{Exciton-polariton spin switches}.
\newblock \emph{\bibinfo{journal}{Nat. Photon.}} \textbf{\bibinfo{volume}{4}},
  \bibinfo{pages}{361--366} (\bibinfo{year}{2010}).

\bibitem{Ballarini2013}
\bibinfo{author}{Ballarini, D.} \emph{et~al.}
\newblock \bibinfo{title}{All-optical polariton transistor}.
\newblock \emph{\bibinfo{journal}{Nat. Comm.}} \textbf{\bibinfo{volume}{4}},
  \bibinfo{pages}{1778} (\bibinfo{year}{2013}).

\bibitem{Steger2013}
\bibinfo{author}{Steger, M.} \emph{et~al.}
\newblock \bibinfo{title}{Long-range ballistic motion and coherent flow of
  long-lifetime polaritons}.
\newblock \emph{\bibinfo{journal}{Phys. Rev. B}} \textbf{\bibinfo{volume}{88}},
  \bibinfo{pages}{235314} (\bibinfo{year}{2013}).

\bibitem{Steger2015}
\bibinfo{author}{Steger, M.}, \bibinfo{author}{Gautham, C.},
  \bibinfo{author}{Snoke, D.~W.}, \bibinfo{author}{Pfeiffer, L.} \&
  \bibinfo{author}{West, K.}
\newblock \bibinfo{title}{Slow reflection and two-photon generation of
  microcavity exciton-polaritons}.
\newblock \emph{\bibinfo{journal}{Optica}} \textbf{\bibinfo{volume}{2}},
  \bibinfo{pages}{1--5} (\bibinfo{year}{2015}).

\bibitem{Zhugayevych2015}
\bibinfo{author}{Zhugayevych, A.} \& \bibinfo{author}{Tretiak, S.}
\newblock \bibinfo{title}{Theoretical description of structural and electronic
  properties of organic photovoltaic materials}.
\newblock \emph{\bibinfo{journal}{Ann. Rev. Phys. Chem.}}
  \textbf{\bibinfo{volume}{66}}, \bibinfo{pages}{305} (\bibinfo{year}{2015}).

\bibitem{Snoke2012}
\bibinfo{author}{Snoke, D.}
\newblock \bibinfo{title}{Polariton condensation and lasing}.
\newblock In \bibinfo{editor}{Timofeev, V.} \& \bibinfo{editor}{Sanvitto, D.}
  (eds.) \emph{\bibinfo{booktitle}{Exciton-Polaritons in Microcavities}}, vol.
  \bibinfo{volume}{172} of \emph{\bibinfo{series}{Springer Series in Solid
  State Sciences}} (\bibinfo{publisher}{Springer}, \bibinfo{year}{2012}).

\bibitem{Tassone1999}
\bibinfo{author}{Tassone, F.} \& \bibinfo{author}{Yamamoto, Y.}
\newblock \bibinfo{title}{Exciton-exciton scattering dynamics in a
  semiconductor microcavity and stimulated scattering into polaritons}.
\newblock \emph{\bibinfo{journal}{Phys. Rev. B}} \textbf{\bibinfo{volume}{59}},
  \bibinfo{pages}{10830--10842} (\bibinfo{year}{1999}).

\bibitem{Prokofev2001}
\bibinfo{author}{Prokof'ev, N.}, \bibinfo{author}{Ruebenacker, O.} \&
  \bibinfo{author}{Svistunov, B.}
\newblock \bibinfo{title}{Critical point of a weakly interacting
  two-dimensional bose gas}.
\newblock \emph{\bibinfo{journal}{Phys. Rev. Lett.}}
  \textbf{\bibinfo{volume}{87}}, \bibinfo{pages}{270402}
  (\bibinfo{year}{2001}).

\bibitem{Prokofev2002}
\bibinfo{author}{Prokof'ev, N.} \& \bibinfo{author}{Svistunov, B.}
\newblock \bibinfo{title}{Two-dimensional weakly interacting bose gas in the
  fluctuation region}.
\newblock \emph{\bibinfo{journal}{Phys. Rev. A}} \textbf{\bibinfo{volume}{66}},
  \bibinfo{pages}{043608} (\bibinfo{year}{2002}).

\bibitem{Anton2015}
\bibinfo{author}{Ant\'on, C.} \emph{et~al.}
\newblock \bibinfo{title}{Optical control of spin textures in
  quasi-one-dimensional polariton condensates}.
\newblock \emph{\bibinfo{journal}{Phys. Rev. B}} \textbf{\bibinfo{volume}{91}},
  \bibinfo{pages}{075305} (\bibinfo{year}{2015}).

\bibitem{Sich2014}
\bibinfo{author}{Sich, M.} \emph{et~al.}
\newblock \bibinfo{title}{Effects of spin-dependent interactions on
  polarization of bright polariton solitons}.
\newblock \emph{\bibinfo{journal}{Phys. Rev. Lett.}}
  \textbf{\bibinfo{volume}{112}}, \bibinfo{pages}{046403}
  (\bibinfo{year}{2014}).

\bibitem{Sekretenko2013}
\bibinfo{author}{Sekretenko, A.~V.}, \bibinfo{author}{Gavrilov, S.~S.} \&
  \bibinfo{author}{Kulakovskii, V.~D.}
\newblock \bibinfo{title}{Polariton-polariton interactions in microcavities
  under a resonant 10 to 100 picosecond pulse excitation}.
\newblock \emph{\bibinfo{journal}{Phys. Rev. B}} \textbf{\bibinfo{volume}{88}},
  \bibinfo{pages}{195302} (\bibinfo{year}{2013}).

\bibitem{Takemura2014_2}
\bibinfo{author}{Takemura, N.}, \bibinfo{author}{Trebaol, S.},
  \bibinfo{author}{Wouters, M.}, \bibinfo{author}{Portella-Oberli, M.~T.} \&
  \bibinfo{author}{Deveaud-Pledran, B.}
\newblock \bibinfo{title}{Heterodyne spectroscopy of polariton spinor
  interactions}.
\newblock \emph{\bibinfo{journal}{Phys. Rev. B}} \textbf{\bibinfo{volume}{90}},
  \bibinfo{pages}{195307} (\bibinfo{year}{2014}).

\bibitem{Wertz2010}
\bibinfo{author}{Wertz, E.} \emph{et~al.}
\newblock \bibinfo{title}{Spontaneous formation and optical manipulation of
  extended polariton condensates}.
\newblock \emph{\bibinfo{journal}{Nat. Phys.}} \textbf{\bibinfo{volume}{6}},
  \bibinfo{pages}{860--864} (\bibinfo{year}{2010}).

\bibitem{Tosi2012_2}
\bibinfo{author}{Tosi, G.} \emph{et~al.}
\newblock \bibinfo{title}{Sculpting oscillators with light within a nonlinear
  quantum fluid}.
\newblock \emph{\bibinfo{journal}{Nat. Phys.}} \textbf{\bibinfo{volume}{8}},
  \bibinfo{pages}{190--194} (\bibinfo{year}{2012}).

\bibitem{Cristofolini2013}
\bibinfo{author}{Cristofolini, P.} \emph{et~al.}
\newblock \bibinfo{title}{Optical superfluid phase transitions and trapping of
  polariton condensates}.
\newblock \emph{\bibinfo{journal}{Phys. Rev. Lett.}}
  \textbf{\bibinfo{volume}{110}}, \bibinfo{pages}{186403}
  (\bibinfo{year}{2013}).

\bibitem{Askitopoulos2013}
\bibinfo{author}{Askitopoulos, A.} \emph{et~al.}
\newblock \bibinfo{title}{Polariton condensation in an optically induced
  two-dimensional potential}.
\newblock \emph{\bibinfo{journal}{Phys. Rev. B}} \textbf{\bibinfo{volume}{88}},
  \bibinfo{pages}{041308} (\bibinfo{year}{2013}).

\bibitem{Askitopoulos2015}
\bibinfo{author}{Askitopoulos, A.} \emph{et~al.}
\newblock \bibinfo{title}{Robust platform for engineering pure-quantum-state
  transitions in polariton condensates}.
\newblock \emph{\bibinfo{journal}{Phys. Rev. B}} \textbf{\bibinfo{volume}{92}}
  (\bibinfo{year}{2015}).

\bibitem{phase_boundaries}
\bibinfo{author}{Sun, Y.} \emph{et~al.}
\newblock \bibinfo{title}{Phase boundaries of multi-mode and single-mode
  quantum states in exciton-polariton condensates}.
\newblock \bibinfo{note}{In preparation}.

\bibitem{Nelsen2013}
\bibinfo{author}{Nelsen, B.} \emph{et~al.}
\newblock \bibinfo{title}{Dissipationless flow and sharp threshold of a
  polariton condensate with long lifetime}.
\newblock \emph{\bibinfo{journal}{Phys. Rev. X}} \textbf{\bibinfo{volume}{3}},
  \bibinfo{pages}{041015} (\bibinfo{year}{2013}).

\bibitem{Zimmermann2006}
\bibinfo{author}{Zimmermann, R.}
\newblock \bibinfo{title}{Dynamical t-matrix theory for high-density excitons
  in coupled quantum wells}.
\newblock \emph{\bibinfo{journal}{Phys. Stat. Sol. (b)}}
  \textbf{\bibinfo{volume}{243}}, \bibinfo{pages}{2358--2362}
  (\bibinfo{year}{2006}).

\bibitem{saturation}
\bibinfo{author}{Ciuti, C.}, \bibinfo{author}{Savona, V.} \&
  \bibinfo{author}{Quattropani, A.}
\newblock \bibinfo{title}{Role of the exchange of carriers in elastic
  exciton-exciton scattering in quantum wells}.
\newblock \emph{\bibinfo{journal}{Phys. Rev. B}} \textbf{\bibinfo{volume}{58}},
  \bibinfo{pages}{7926} (\bibinfo{year}{1998}).

\bibitem{pavlovic}
\bibinfo{author}{Pavlovic, G.}
\newblock \emph{\bibinfo{title}{Exciton-polaritons in low dimensional
  structures}}.
\newblock Ph.D. thesis, \bibinfo{school}{Universit\'e Blaise Pascal -
  Clermont-Ferrand} (\bibinfo{year}{2010}).

\bibitem{Takemura2014}
\bibinfo{author}{Takemura, N.}, \bibinfo{author}{Trebaol, S.},
  \bibinfo{author}{Wouters, M.}, \bibinfo{author}{Portella-Oberli, M.~T.} \&
  \bibinfo{author}{Deveaud, B.}
\newblock \bibinfo{title}{Polaritonic feshbach resonance}.
\newblock \emph{\bibinfo{journal}{Nat. Phys.}} \textbf{\bibinfo{volume}{10}},
  \bibinfo{pages}{500--504} (\bibinfo{year}{2014}).

\bibitem{Marchetti2006}
\bibinfo{author}{Marchetti, F.~M.}, \bibinfo{author}{Keeling, J.},
  \bibinfo{author}{Szyma\ifmmode~\acute{n}\else \'{n}\fi{}ska, M.~H.} \&
  \bibinfo{author}{Littlewood, P.~B.}
\newblock \bibinfo{title}{Thermodynamics and excitations of condensed
  polaritons in disordered microcavities}.
\newblock \emph{\bibinfo{journal}{Phys. Rev. Lett.}}
  \textbf{\bibinfo{volume}{96}}, \bibinfo{pages}{066405}
  (\bibinfo{year}{2006}).

\bibitem{Marchetti2007}
\bibinfo{author}{Marchetti, F.~M.}, \bibinfo{author}{Keeling, J.},
  \bibinfo{author}{Szyma\ifmmode~\acute{n}\else \'{n}\fi{}ska, M.~H.} \&
  \bibinfo{author}{Littlewood, P.~B.}
\newblock \bibinfo{title}{Absorption, photoluminescence, and resonant rayleigh
  scattering probes of condensed microcavity polaritons}.
\newblock \emph{\bibinfo{journal}{Phys. Rev. B}} \textbf{\bibinfo{volume}{76}},
  \bibinfo{pages}{115326} (\bibinfo{year}{2007}).

\bibitem{Taylor2015}
\bibinfo{author}{Taylor, J.}
\newblock \bibinfo{howpublished}{private communication}.

\bibitem{Kavokin2007}
\bibinfo{author}{Kavokin, A.}, \bibinfo{author}{Baumberg, J.~J.},
  \bibinfo{author}{Malpuech, G.} \& \bibinfo{author}{Laussy, F.~P.}
\newblock \emph{\bibinfo{title}{Microcavities}} (\bibinfo{publisher}{Oxford
  Science Publications}, \bibinfo{year}{2007}).

\bibitem{Deng2010}
\bibinfo{author}{Deng, H.}
\newblock \bibinfo{title}{Exciton-polariton bose-einstein condensation}.
\newblock \emph{\bibinfo{journal}{Rev. Mod. Phys.}}
  \textbf{\bibinfo{volume}{82}}, \bibinfo{pages}{1489--1537}
  (\bibinfo{year}{2010}).

\bibitem{Snoke2010}
\bibinfo{author}{Snoke, D.~W.} \& \bibinfo{author}{Littlewood, P.}
\newblock \bibinfo{title}{Polariton condensates}.
\newblock \emph{\bibinfo{journal}{Physics Today}}
  \textbf{\bibinfo{volume}{63}}, \bibinfo{pages}{42--47}
  (\bibinfo{year}{2010}).

\bibitem{Carusotto2013}
\bibinfo{author}{Carusotto, I.} \& \bibinfo{author}{Cuiti, C.}
\newblock \bibinfo{title}{Quantum fluids of light}.
\newblock \emph{\bibinfo{journal}{Rev. Mod. Phys.}}
  \textbf{\bibinfo{volume}{85}}, \bibinfo{pages}{299--366}
  (\bibinfo{year}{2013}).

\bibitem{Hopfield1958}
\bibinfo{author}{Hopfield, J.~J.}
\newblock \bibinfo{title}{Theory of the contribution of excitons to the complex
  dielectric constant of crystals}.
\newblock \emph{\bibinfo{journal}{Phys. Rev.}} \textbf{\bibinfo{volume}{112}},
  \bibinfo{pages}{1555--1567} (\bibinfo{year}{1958}).

\bibitem{thermalization}
\bibinfo{author}{Sun, Y.} \emph{et~al.}
\newblock \bibinfo{title}{Exciton-polariton bose-einstein condensates at
  equilibrium}.
\newblock \bibinfo{note}{Unpublished}.

\bibitem{Ramsteiner1997}
\bibinfo{author}{Ramsteiner, M.} \emph{et~al.}
\newblock \bibinfo{title}{Influence of composition fluctuations in al(ga)as
  barriers on the exciton localization in thin gaas quantum wells}.
\newblock \emph{\bibinfo{journal}{Phys. Rev. B}} \textbf{\bibinfo{volume}{55}},
  \bibinfo{pages}{5239--5242} (\bibinfo{year}{1997}).

\bibitem{Zhao2002}
\bibinfo{author}{Zhao, H.}, \bibinfo{author}{Moehl, S.},
  \bibinfo{author}{Wachter, S.} \& \bibinfo{author}{Kalt, H.}
\newblock \bibinfo{title}{Hot exciton transport in znse quantum wells}.
\newblock \emph{\bibinfo{journal}{Appl. Phys. Lett.}}
  \textbf{\bibinfo{volume}{80}}, \bibinfo{pages}{1391--1393}
  (\bibinfo{year}{2002}).

\bibitem{Akselrod2010}
\bibinfo{author}{Akselrod, G.~M.}, \bibinfo{author}{Tischler, Y.~R.},
  \bibinfo{author}{Young, E.~R.}, \bibinfo{author}{Nocera, D.~G.} \&
  \bibinfo{author}{Bulovic, V.}
\newblock \bibinfo{title}{Exciton-exciton annihilation in organic polariton
  microcavities}.
\newblock \emph{\bibinfo{journal}{Phys. Rev. B}} \textbf{\bibinfo{volume}{82}}
  (\bibinfo{year}{2010}).

\bibitem{Akselrod2014}
\bibinfo{author}{Akselrod, G.~M.} \emph{et~al.}
\newblock \bibinfo{title}{Visualization of exciton transport in ordered and
  disordered molecular solids}.
\newblock \emph{\bibinfo{journal}{Nat. Comm.}} \textbf{\bibinfo{volume}{5}},
  \bibinfo{pages}{1--8} (\bibinfo{year}{2014}).

\end{thebibliography}

\vspace{10pt}
\noindent\textbf{Acknowledgements} \\ We thank A. Daley, and D. Pekker for fruitful discussions and J. Beaumarriage for assistance in the calibration of the detuning map of the sample.  Y.S. and Y.Y. and K.A.N were supported as part of the Center for Excitonics, an Energy Frontier Research Center funded by the US Department of Energy, Office of Science, Office of Basic Energy Sciences under Award Number DE-SC0001088. M.S., G.L. and D.W.S. were  supported  by  the  National  Science  Foundation under grants PHY-1205762 and DMR-1104383. L.N.P. and K.W. were funded by the Gordon and Betty Moore Foundation through the EPiQS initiative Grant GBMF4420, and by the National Science Foundation MRSEC Grant DMR-1420541.\\
\\
\noindent\textbf{Author contributions} \\ Y.S., D.W.S, and K.A.N. designed the experiments; Y.S. and Y.Y. performed the experiment; Y.S. and D.W.S. analyzed the data; Y.S., Y.Y. and M.S. calibrated the detuning map of the sample; L.P.N and K.W. grew the microcavity structure; all the authors participated to the result discussion and manuscript preparation.\\
\\
\noindent\textbf{Additional information} \\Supplementary information is available in the online version of the paper. Reprints and permissions information is available online at www.nature.com/reprints. Correspondence and requests for materials should be addressed to Y.S. at \href{mailto:ybsun@mit.edu}{ybsun@mit.edu} or to  D.W.S. at \href{mailto:snoke@pitt.edu}{snoke@pitt.edu}.\\
\\
\noindent\textbf{Competing financial interest} \\The authors declare no competing financial interests.

\newpage
\subsection*{Supplementary Information for ``Polaritons are Not Weakly Interacting: Direct Measurement of the Polariton-Polariton Interaction Strength''}
\begin{figure}[htbp]
\centering
  \includegraphics[width=0.85\textwidth]{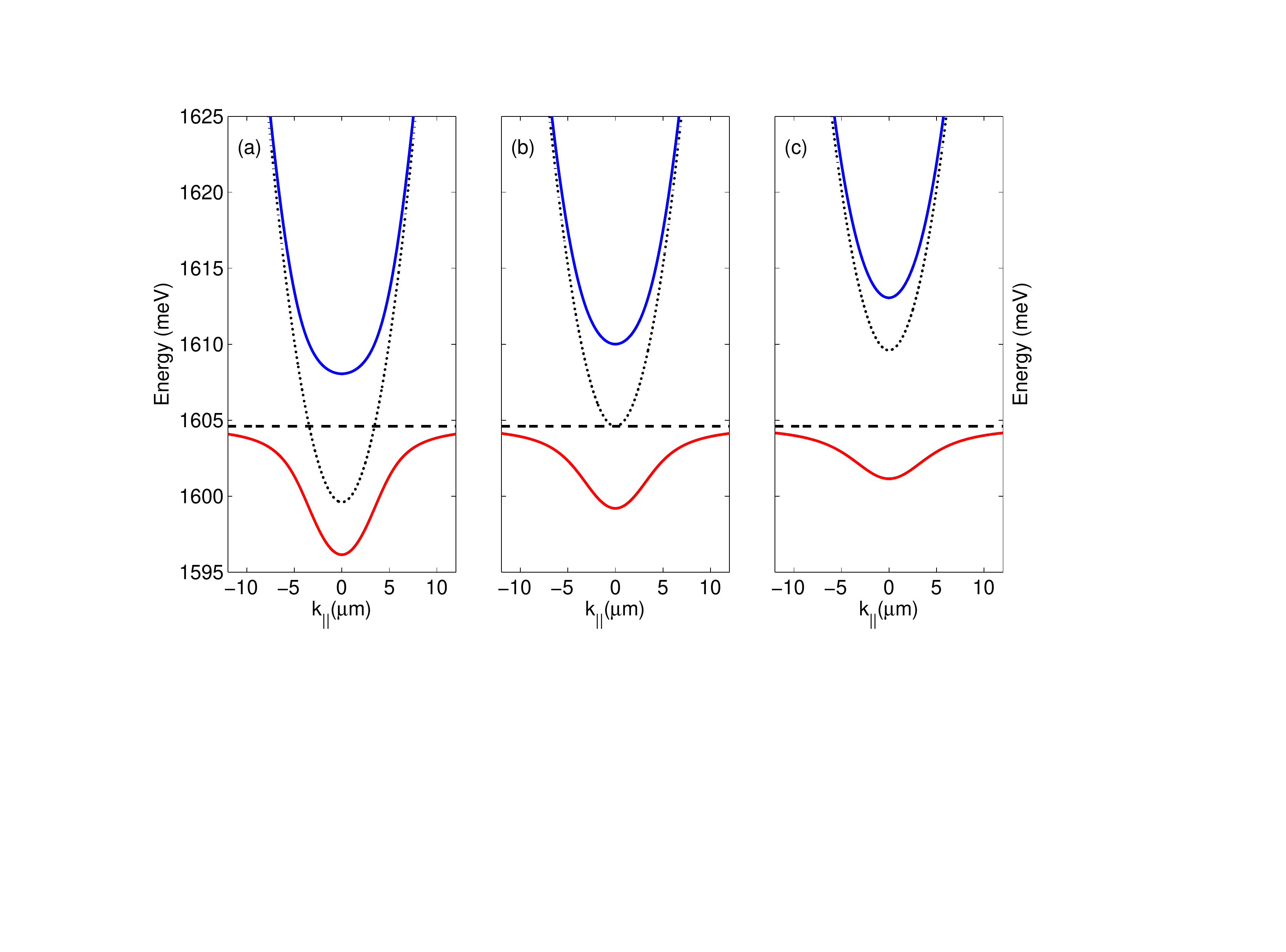}
   \caption{ (color online) Dispersion curves of polariton at three representative cavity detunings (a) $\delta = -5$ meV. (b) $\delta = 0$ meV. (c) $\delta = 5$ meV. The dotted line shows the confined cavity mode, and the dashed line shows the bare exciton mode. The blue and red solid lines indicate the upper polariton (LP) and lower polariton (UP) branches, respectively, arising from the strong coupling between corresponding cavity modes and exciton modes. Calibrated sample parameters were used in the calculations.}
\label{energy_dispersion}
\end{figure}

\noindent\textbf{\textrm{Background on exciton-polaritons in semiconductor microcavities}}.\hspace{10pt}  Exciton-polaritons are formed in semiconductor microcavities through the strong coupling between optical modes of the microcavity and exciton transitions of material embedded inside the microcavity \cite{Kavokin2007, Deng2010, Snoke2010, Carusotto2013}. For the case of a single microcavity mode and a single exciton transition, two polariton modes, the upper and lower polaritons, are formed with the energies of the two polariton modes, $E_{LP}(k_{||})$ and $E_{UP}(k_{||})$, given by:
\begin{equation}
E_{LP/UP}(k_{||}) = \frac{1}{2}\left[E_{X}({k_{||}}) + E_{C}(k_{||}) \mp \sqrt{\Omega^2 + \delta^2(k_{||})}\right]
\label{eq:Epolariton}
\end{equation}
where $k_{||}$ is the wave vector in the plane perpendicular to the microcavity confinement direction, $E_X(k_{||})$ is the energy of the exciton transition, $E_C(k_{||})$ is the energy of the cavity mode, $\delta(k_{||})$ is the detuning energy defined as $\delta(k_{||})=E_C(k_{||})-E_X(k_{||})$, and $\Omega$ is the strength of radiative coupling between the exciton and cavity field, also known as full Rabi splitting energy. The confinement of light gives the cavity mode a parabolic dispersion in the plane perpendicular to the confinement direction: $E_C=\hbar^2k_{||}^2/2m_C$, where $m_C$ is the effective mass of the cavity field. This effective mass is typically $10^{-4}$ times lighter than the vacuum electron mass, and about $10^{-3}$ times less than an exciton in a GaAs quantum well structure, so that $E_X(k_{||})$ is essentially constant with $k_{||}$. The energies $E_X(k_{||}),E_C(k_{||}),E_{LP}(k_{||})$, and $E_{UP}(k_{||})$ as given in Fig.~\ref{energy_dispersion} for three different values of $\delta(k_{||}=0)$. The energies were calculated using (\ref{eq:Epolariton}) and parameters matching the sample structure used in the experiments: $\Omega = 10.84$ meV and $E_X(0)=1604.6$ meV. 

\begin{figure}[htbp]
\centering
  \includegraphics[width=0.50\textwidth]{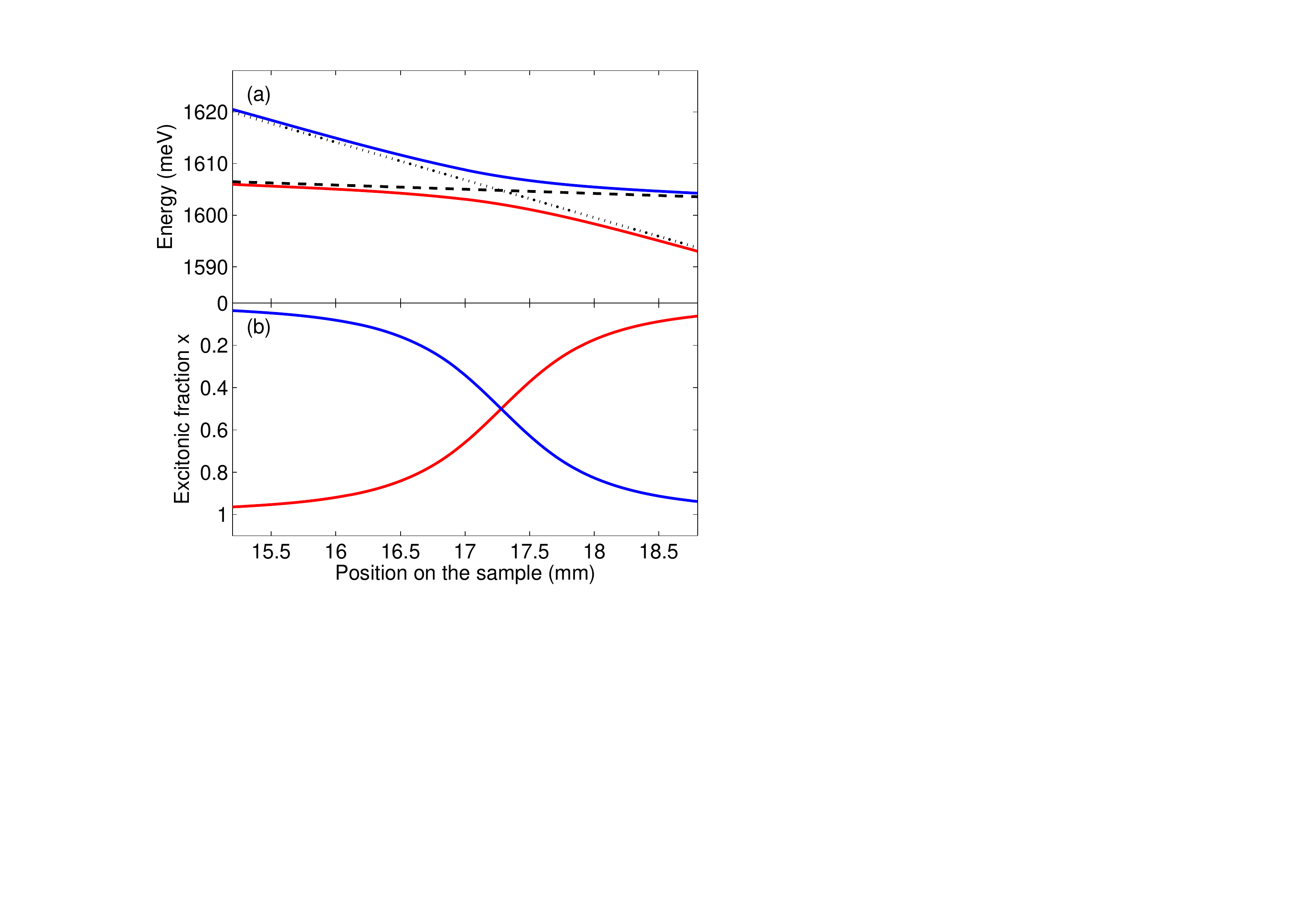}
   \caption{ (color online) (a) The calculated upper polariton (blue line) and lower polariton (red line) energies at different positions of the sample. The dashed lines indicate the exciton energies, and the dotted line shows the cavity energies. (b) Excitonic fractions of upper polaritons (blue line) and  lower polaritons (red line) at different sample positions.}
\label{spatial_dispersion}
\end{figure}

The length of the cavity increases monotonically along one direction of the QW plane so that the energy of the cavity mode can be tuned relative to the exciton resonance energy, as shown in Fig.~\ref{spatial_dispersion}, allowing us to experimentally tune $\delta(k_{||}=0)$. The energies of all modes in Fig.~\ref{energy_dispersion} are plotted as a function of $k_{||}$, the in-plane wave vector. As can be seen in this figure, $E_X(k_{||})$ is essentially constant with respect to $k_{||}$ while $E_C(k_{||})$ is parabolic.

The polariton modes are linear superpositions of the exciton and microcavity photon modes. The lower polariton and upper polariton operators, $\hat{P}_{k_{||}}$ and $\hat{Q}_{k_{||}}$, respectively, can be written in terms of exciton and cavity operators, $\hat{a}_{k_{||}}$ and $\hat{b}_{k_{||}}$:
\begin{eqnarray}
\hat{P}_{k_{||}} &= X(k_{||})\hat{a}_{k_{||}}+C(k_{||})\hat{b}_{k_{||}k_{||}}\\
\hat{Q}_{k_{||}} &=-C(k_{||})\hat{a}_{k_{||}}+X(k_{||})\hat{b}_{k_{||}}.
\end{eqnarray}
The coefficients, $X(k_{||})$ and $C(k_{||})$, are called the exciton and cavity Hopfield coefficients \cite{Hopfield1958} and are given by 
\begin{eqnarray}
|X(k_{||})|^2 &= \frac{1}{2}\left(1+\frac{\delta(k_{||})}{\sqrt{\delta^2(k_{||})+\Omega^2}}\right)\\
|C(k_{||})|^2 &= \frac{1}{2}\left(1-\frac{\delta(k_{||})}{\sqrt{\delta^2(k_{||}k_{||})+\Omega^2}}\right)
\end{eqnarray}

The characteristics of the polariton modes are determined by the coefficients, which depend on $\delta(k_{||})$. The lower polariton is more photon-like and the upper polariton is more exciton-like for $\delta(k_{||})<0$, and the lower polariton is more exciton-like and the upper polariton is more photon like $\delta(k_{||})>0$. Due to the wedge in the cavity thickness, we can easily tune the excitonic fraction $|X(k_{||})|^2$ of lower polaritons by moving the excitation spot at different positions, as shown in Fig.~\ref{spatial_dispersion}b, where we plot $|X(k_{||}=0)|^2$ at different positions on the sample. As seen in Fig.~\ref{energy_dispersion} and Fig.~\ref{spatial_dispersion}, the energies and shapes of the polariton dispersion curves depend strongly on $\delta$:
positive detuning results in lower polaritons that are more exciton-like, with a heavier effective mass and stronger interactions with phonons and other carriers, while negative detuning results in lower polaritons that are more photon-like, with a smaller lower polariton mass and weaker interactions with phonons and other carriers.

\begin{figure}[htbp]
\centering
  \includegraphics[width=0.50\textwidth]{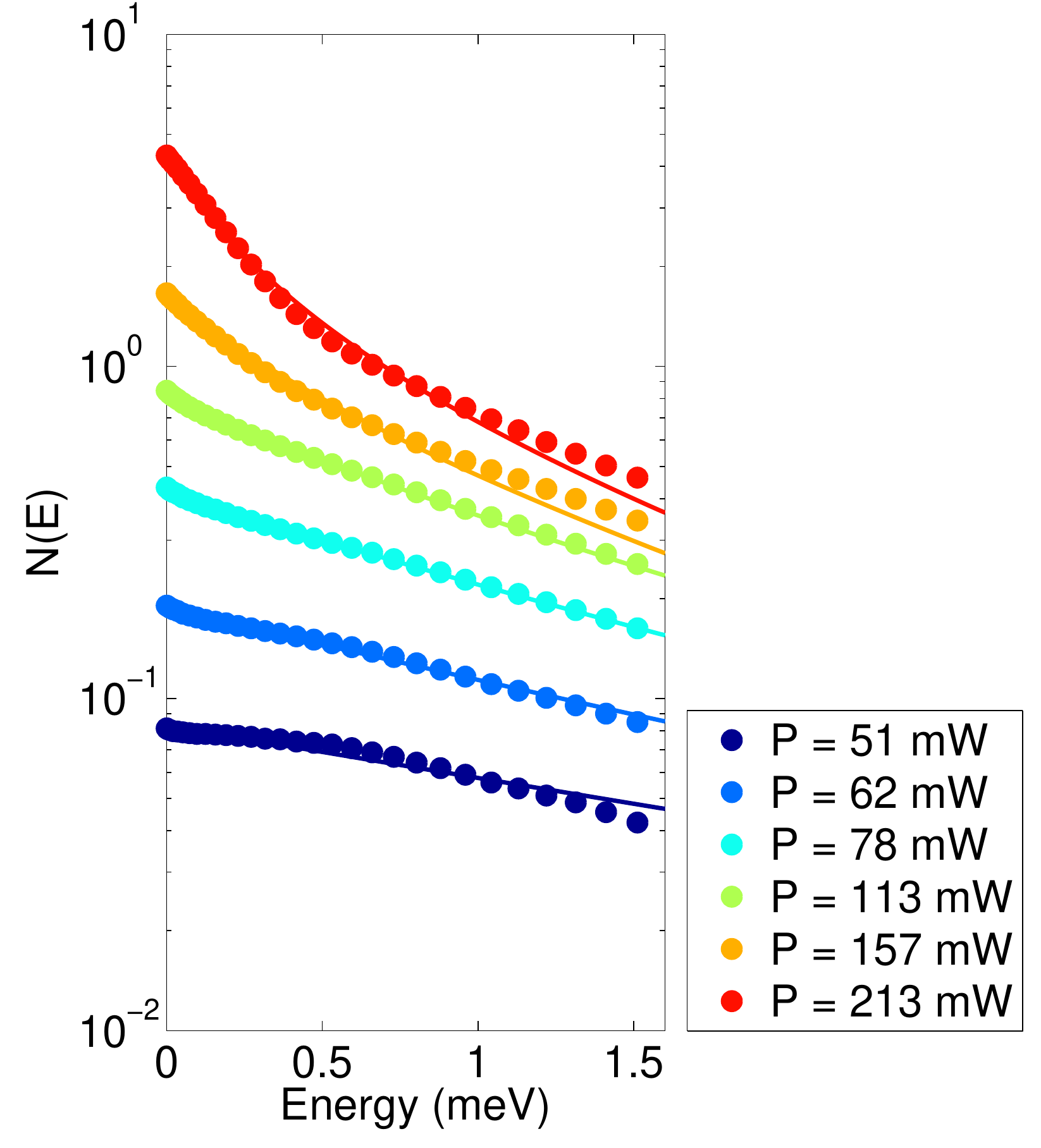}
   \caption{ (color online). Energy distributions of polaritons at different excitation densities (solid dots) and fits to Bose-Einstein distributions at high densities (solid lines) in a ring with a diameter of 72 $\mu$m at the cavity detuning of $\delta = 4.31$ meV, corresponding to the case in Fig.~5 in the main text. The fitted temperatures $T$ and reduced chemical potentials $\mu/k_BT$ are 24.9 K, 20.6 K, 18.9 K, 17.9 K, 15.3 K, 15.2 K and $-2.6,  -1.8,  -1.2, -0.8, -0.5$ and $-0.2$, from low densities to high densities, respectively.}
\label{energy_distribution}
\end{figure}
\vspace{.5cm}
\noindent\textbf{\textrm{Energy distributions of polaritons at different excitation densities}}.\hspace{10pt} Due to the high $Q$ of the microcavity structure used in this experiment, polaritons can well equilibrate before coupling out of the cavity.  Fig.~\ref{energy_distribution}, corresponding to Fig.~5 in the main text, shows data for $N(E)$ for various pumps powers, deduced from energy-resolved $k$-space data using the calibration factor discussed below. The solid lines are fits to a Bose-Einstein distribution $N(E) = 1/(e^{(E-\mu)/k_BT}-1)$.  At very low density, the distribution is not well thermalized, while at intermediate density, the polaritons fit to a Maxwell-Boltzmann distribution, which corresponds to a straight line on this semilog plot. In this range of density, the fit value of the chemical potential $\mu$ just gives an overall multiplier, as $N(E) \simeq e^{\mu/k_BT}e^{-E/k_BT}$.  
At higher density, there is an upturn in the low-energy states near the ground state, and the distributions fit the equilibrium Bose-Einstein distribution. In this regime not only the overall multiplier but also the curvature of the distribution must be fit by the parameter $\mu$. As seen in Figure~\ref{energy_distribution}, in the high-density regime, the Bose-Einstein equilibrium fits get both the curvature and the absolute magnitude of $N(E)$ right, with small (10\%) deviations for high-energy states. Alternatively, corrections for the effects of interactions could be made for the low-energy states. This will be the subject of another publication \cite{thermalization}. 

\begin{figure}[htbp]
\centering
  \includegraphics[width=0.99\textwidth]{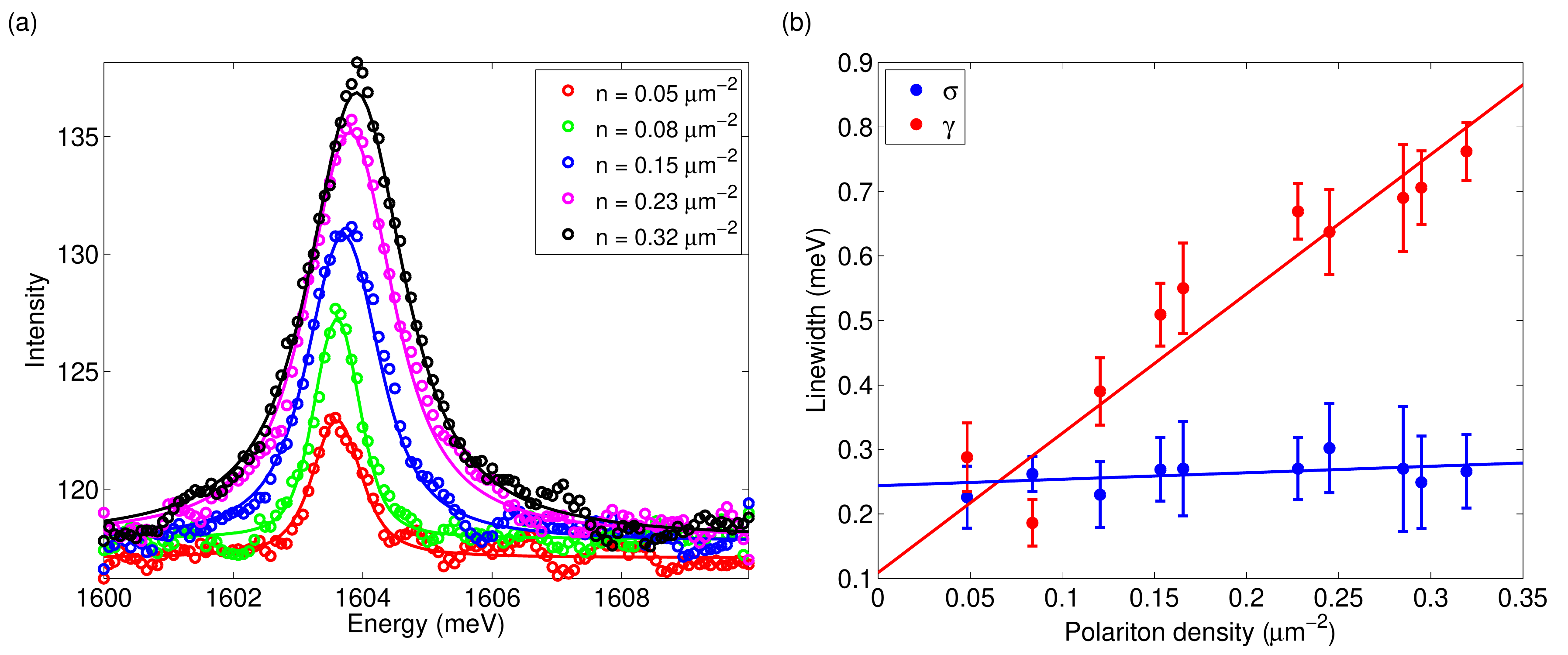}
   \caption{ (color online). (a) Energy emission profiles at $k_{||}=0$ at different excitation densities for the detuning of $\delta = 10.11$ meV with a ring radius of 85 $\mu$m corresponding to Fig.~2 in the main text. The dots are five-point running average of the raw data, and the solid lines are the fits to the Voigt profile. (b) Gaussian linewidth $\sigma$ and Lorentzian linewidth $\gamma$ from the fitting in (a) at different excitation densities. The solid lines are best fits to a linear dependence using weighted least-squares estimates. }
\label{emission_profile}
\end{figure}

\vspace{.5cm}
\noindent\textbf{\textrm{Emission profile at $k_{||}=0$ at different intensities.}} \hspace{10pt} The emission profile at $k_{||}=0$ at different excitation densities for the detuning of $\delta = 10.11$ meV, corresponding to the case in Fig.~2 in the main text, is shown in Fig.~\ref{emission_profile}. The background signal is not subtracted, and the solid lines are the fit to a Voigt lineshape, given by
\begin{align}
V(E;\sigma,\gamma,E_0,A,c) &=  A\int_{-\infty}^\infty G(E'; \sigma,  E_0) L(E-E';\gamma, E_0) dE' +c
\end{align}
where $G(E;\sigma,E_0)$ is the Gaussian profile
\begin{align}
G(E;\sigma,E_0) &= \frac{e^{-(E-E_0)^2/2\sigma^2}}{\sigma\sqrt{2\pi}}
\end{align}
and $L(E;\gamma,E_0)$ is the Lorentzian profile
\begin{align}
L(E;\gamma, E_0) &= \frac{1}{\pi}\frac{\gamma}{(E-E_0)^2+\gamma^2};
\end{align}
$A$ is the amplitude of the profile, and $c$ is the offset. The superimposed solid lines are best fits to a linear dependence. As can be seen, the homogeneous linewidth $\gamma$ shows strong density effect because the increase in the polariton density will lead to more frequent collisions. However, the inhomogeneous linewidth $\sigma$ remains approximately the same in the range of densities used in the measurements.

\vspace{.5cm}
\noindent\textbf{\textrm{Determination of the efficiency factor}}.\hspace{10pt} While the overall conversion factor was allowed to vary to fit the absolute scale of the polariton numbers, we had a way to check that its value was reasonable by determining the number of photons emitted from the sample. A flat mirror was placed at the sample position and aligned to the focal plane of the objective lens that collects the photoluminescence in the typical experiment. A collimated laser beam at 780 nm was sent to the excitation path and was focused by a microscope objective. The back-reflected light was collected by the same objective lens, and went though the same collection path in the far-field geometry as being used for the blue shift measurement. 
The overall efficiency of the optical path, including the efficiency of the CCD and optical grating used to disperse the light, is 
\begin{align}
\xi &= \frac{Ch\nu}{P} \label{a2}
\end{align}
where $P$ is the power measured at the sample position, $C$ is the corresponding CCD counts per second, and $h\nu$ is the photon energy of the excitation beam. The quantity $\xi$ can also be estimated by multiplying the efficiency of each optic in the imaging system including the efficiency of the dispersing grating and spectrometer CCD .

We want to get the distribution $N(E)$ from the spectral data $I\left(E,k\right)$, where $I(E,k)$ is the number of counts per second on the CCD at specific energy $E$ and momentum $k$ from a plot like that shown in Fig.~2(c) or (d). By integrating the intensity counts $I(E,k)$ along the energy axis, and assigning an energy based on the polariton dispersion $E(k)$ for each $k$ state, we obtain $I(E)$.  To emphasize the discretized nature of the images, we specifically write down $I(E)$ as $I_i$ to indicate the number of counts per second for a given pixel $i$. We multiply the $I_i$ data by 
\begin{align}
\eta &= \frac{1}{\xi M_i} \label{a1}
\end{align}
where $\xi$ is the overall efficiency of the optical path given above, and $M_i$ is the number of $k$-states contained within the pixel $i$ of the Fourier plane imaged formed on the CCD camera.

\begin{figure}[htbp]
\centering
\includegraphics[width=0.75\textwidth]{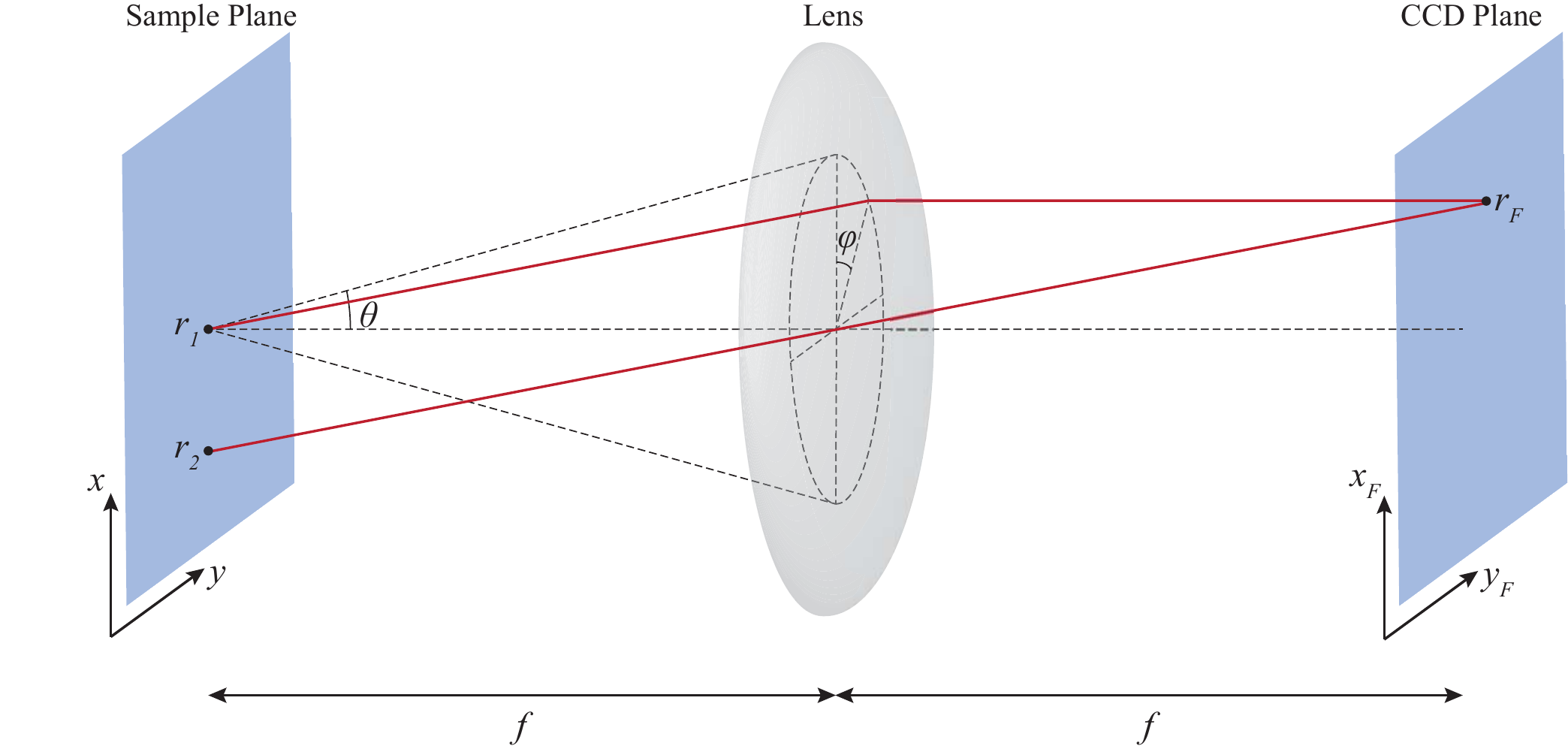}
\caption{Schematic illustration of the far field imaging used in the experiment. $(x,y)$ indicates the sample plane, and $(x_F,y_F)$ are the plane of the spectrometer CCD placed at the Fourier plane of the imaging lens, shown as the gray ellipse. }
\label{optics}
\end{figure}

In order to estimate $M_i$, we calculate the density of states in the plane of spectrometer CCD. A schematic illustration of the optical path is shown in Fig.~\ref{optics}, with $(x,y)$ and $(x_F,y_F)$ indicating the sample plane and the CCD plane at the Fourier plane of the imaging lens. Far-field imaging allows the mapping of rays $(r_1,r_2)$ that emits in the same angle $(\theta,\phi)$  on the sample plane to the same spatial position on the CCD as $r_F$.

Let $(k_x,k_y)$ label the Cartesian coordinate in the reciprocal space at the sample plane, and let $(x_F,y_F)$ label the Cartesian coordinate on the CCD. Define the angle $\theta$ as the emitting angle with respect to the normal of the sample plane, and $\phi$ as the azimuthal angle of the emission at the sample plane, as shown in Fig.~\ref{optics}. Then we have
\begin{align}
k_{||} &= k \sin\theta\\
\tan\phi &= \frac{k_y}{k_x}
\end{align}
where $k_{||}$ and $k$ are the in-plane momentum and the magnitude of the total momentum, with the $x$ and $y$ components being $k_x$ and $k_y$ respectively.  The $k(\theta)$ can be obtained from the dispersion relation as 
\begin{align}
k &= \frac{1}{2\hbar c}\left(E_x+E_c+\frac{\hbar^2}{2m}k_{||}^2-\sqrt{\left(E_x-E_c-\frac{\hbar^2}{2m}k^2_{||}\right)^2+\Omega^2}\right) \label{lpdisp}
\end{align}
where $E_x$ and $E_c$ are the exciton energy and cavity energy at $k_{||}=0$, $\Omega$ is the Rabi energy. Connecting the emission angles to the Cartesian coordinates on the sample plane $(x,y)$ is done using the following equations:
\begin{align}
x &= \gamma f\tan\theta \cos\phi \\
y &= \gamma f\tan\theta \sin\phi.  
\end{align}
Putting all pieces together, yielding                      
\begin{align}
x &= \gamma f\tan\left(\arcsin\frac{k_{||}}{k}\right)\cos\left(\arctan\frac{k_y}{k_x}\right)\\
y &= \gamma f\tan\left(\arcsin\frac{k_{||}}{k}\right)\sin\left(\arctan\frac{k_y}{k_x}\right)
\end{align}
Then the area element will be connected through Jacobian as 
\begin{align}
dxdy &=\frac{\partial(x,y)}{\partial(k_x,k_y)} dk_xdk_y
\end{align}
In order to simplify the calculation of the Jacobian, we approximate the dispersion relation in Eq.~(\ref{lpdisp}) as a quadratic relation given as 
\begin{align}
k &= k_0 + \frac{\hbar^2}{2mc}k_{||}^2
\end{align}
where $k_0 = E_0/\hbar c$, and $E_0$ is the ground state energy at $k_{||}=0$. Working out the Jacobian gives
\begin{align}
\frac{\partial(x,y)}{\partial(k_x,k_y)}  &=  \left|\begin{matrix}\frac{\partial x}{\partial k_x} & \frac{\partial x}{\partial k_y} \\
\frac{\partial y}{\partial k_x} &\frac{\partial y}{\partial k_y}
\end{matrix}\right| = \frac{\gamma^2f^2\left(k^2+akk_{||}\right)}{(k^2-k_{||}^2)^2}
\end{align}
with $a = \hbar/mc$. Therefore,
\begin{align}
dn &= \frac{S}{4\pi^2}dk_xdk_y = \frac{S}{4\pi^2}\frac{(k^2-k^2_{||})^2}{\gamma^2 f^2(k^2+akk_{||})}dxdy
\end{align}
where $S/4\pi^2$ is the density of $k$ states in the reciprocal space, and the density of $k$ states per pixel on the CCD is 
\begin{align}
M_i &= \frac{SA(k^2-k^2_{||})^2}{4\pi^2\gamma^2f^2(k^2+akk_{||})}
\end{align}
with $A$ being the area of one pixel. This factor gives a correction of $\sim$12\% at the edge of our field of view compared to $k_{||}=0$. The total number of polaritons in the field of view can be computed using 
\begin{align}
N_{LP} = \sum_{i} \frac{N_i\tau_i}{\xi M_i} \label{nlp}
\end{align}
where $N_i$ is the counts per second on the CCD in the measurement, and $\tau_i$ is the lifetime of polaritons, which is also angle-dependent.

An independent check of the efficiency factor can be done by calculating the number of polaritons from integrating over the Bose-Einstein distributions as shown in Fig.~\ref{energy_distribution} using 
\begin{align}
N_{LP} &= \frac{gmS}{2\pi \hbar^2}\int_0^{E_{max}} N(E)dE
\end{align} 
where $S$ is area of the field of view, and $E_{max}$ is the upper bound of the energy limit collected by the objective lens. From this way, we compute that the number of polaritons in the field of view at a pumping power of $P = 78$ mW corresponding to dark green curve in Fig.~\ref{energy_distribution} is $N_{LP} = 1058$. 
By using Eq.~(\ref{nlp}), we get a quantity of $\sum N_{LP} =$ 1012, which is less than 5\% different.

\begin{figure}[htbp]
\centering
  \includegraphics[width=0.50\textwidth]{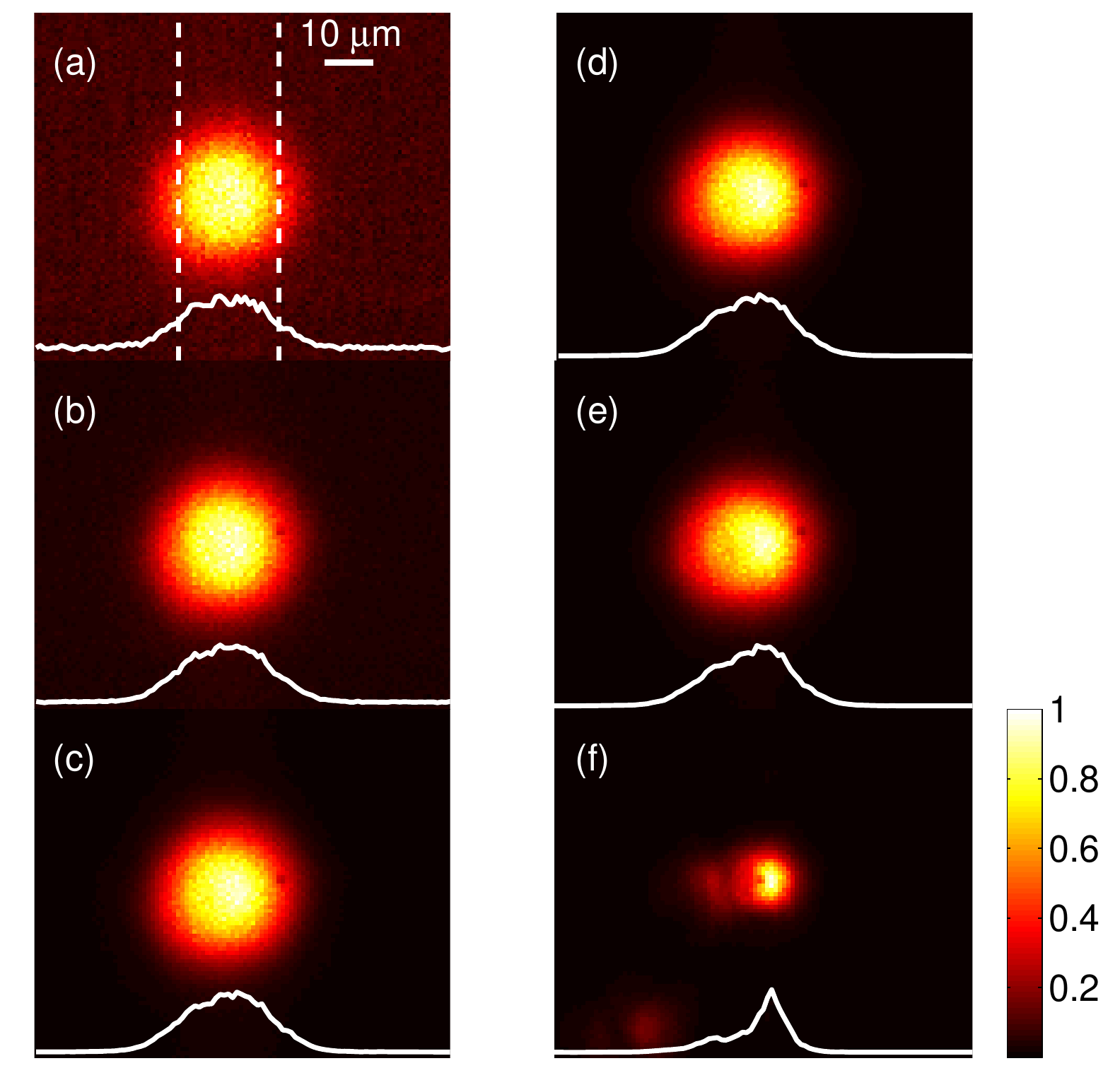}
   \caption{ (color online) (a)-(f) Spatial distributions of polaritons at different excitation densities in an annular trap with a diameter of 56 $\mu$m at a cavity detuning of $\delta = 4.63$ meV. (a) $P= 0.09\, P_{c}$ (b) $P=0.31\, P_c$ (c) $P=P_c$ (d) $P=2.42 \, P_c$ (e) $P=3.26\,P_c$ (f) $P=5.71\, P_c$, with $P_c$ denotes the threshold pumping power. The dashed line indicates the collection region, and the solid line is the cross section of the horizontal slice at the center. The long tail from (f) is from the scattering of the neutral density filter used to attenuate the emission.}
\label{spatial_distribution}
\end{figure}

\vspace{.5cm}
\noindent\textbf{\textrm{Spatial distributions of exciton-polaritons below and above condensation threshold}}.\hspace{10pt} The spatial distributions of the polaritons are nearly homogenous in the field of view from which the data was collected. Fig.~\ref{spatial_distribution} shows the spatial profiles of the polaritons in an annular trap with a diameter of 56 $\mu$m at a detuning of $\delta = 4.63$ meV for a wide range of pumping powers. PL from the barrier was spatially filtered out, otherwise the emission pattern will be dominated by the PL at the pumping region. Within the collection region where the energy shifts were measured in the main text, indicated by the dashed line, the spatial distributions vary by less than 20\% over the field of view at either below the threshold as shown in (a)-(b), or at threshold as shown in (c), or above the threshold as shown in (d)-(e). For rings with larger diameters, the variation within the field of view should be even less. The white solid lines are the edges of the field of view at the center used for the energy shift data. As seen in this figure, the real-space profiles remain similar over a broad range of polariton densities. When the system was pumped very hard, self-trapping of the polariton condensate was observed, as seen in Fig.~\ref{spatial_distribution}f.  

\begin{figure}[htbp]
\centering
  \includegraphics[width=0.5\textwidth]{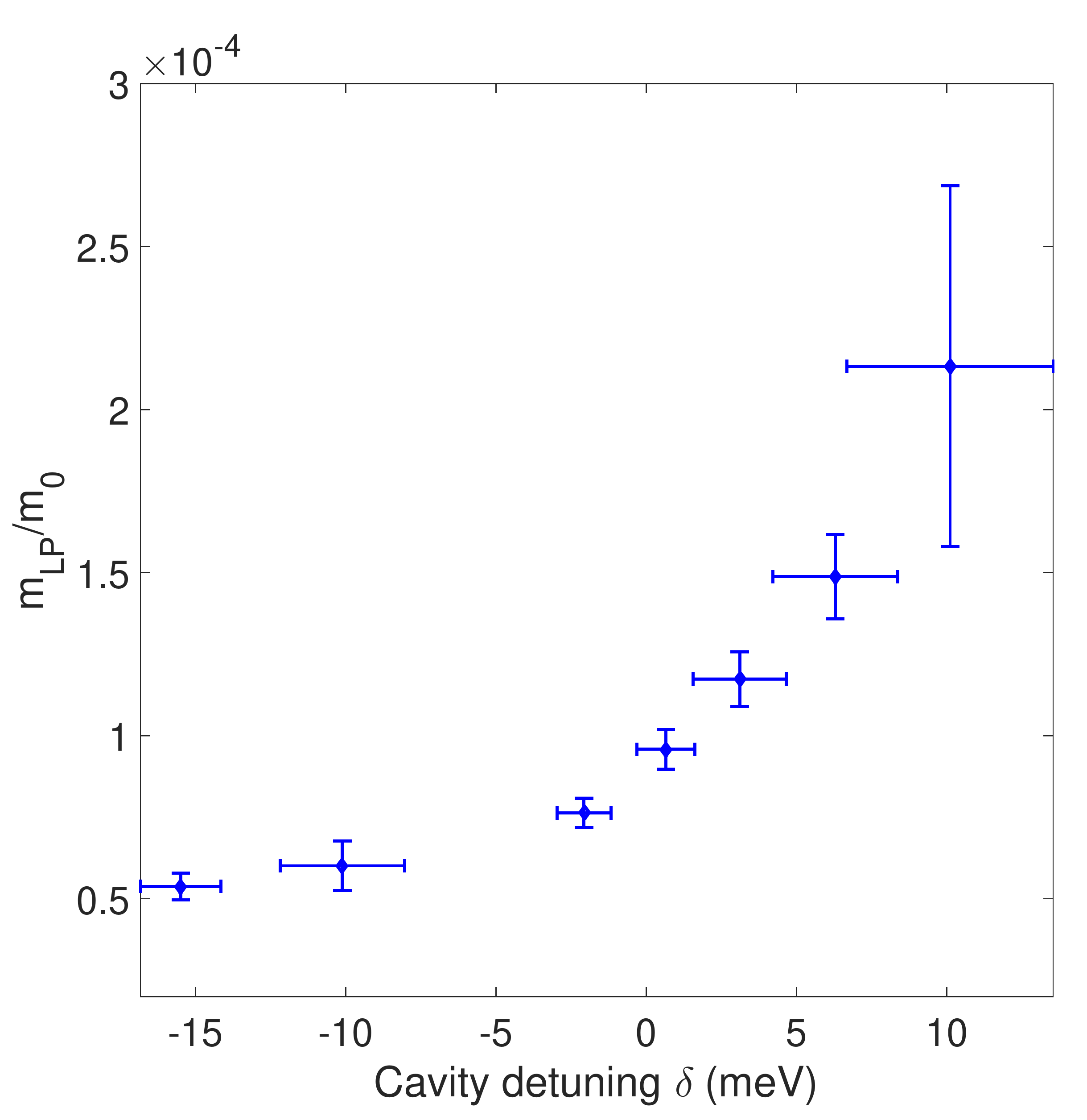}
   \caption{Effective mass of the polaritons as a function of the detuning, where $m_0$ is the vacuum electron mass deduced from the energy dispersions collected in a ring with a radius of 85 $\mu$m, corresponding to that in Fig.~3 in the main text.}
\label{mass}
\end{figure}

\vspace{.5cm}
\noindent\textbf{\textrm{Effective mass of the polaritons}}.\hspace{10pt} Fig.~\ref{mass} shows the effective mass of the polaritons as the detunings used for Fig.~4 of the main text, deduced from the Hopfield coefficients used, which were derived from the detuning, as discussed above. As can be seen, the masses are of four orders of magnitude less than the vacuum electron mass, or three orders of magnitude smaller than the typical exciton mass in GaAs thin quantum wells, in the range of our measurements.

\begin{figure}[htbp]
\centering
  \includegraphics[width=0.5\textwidth]{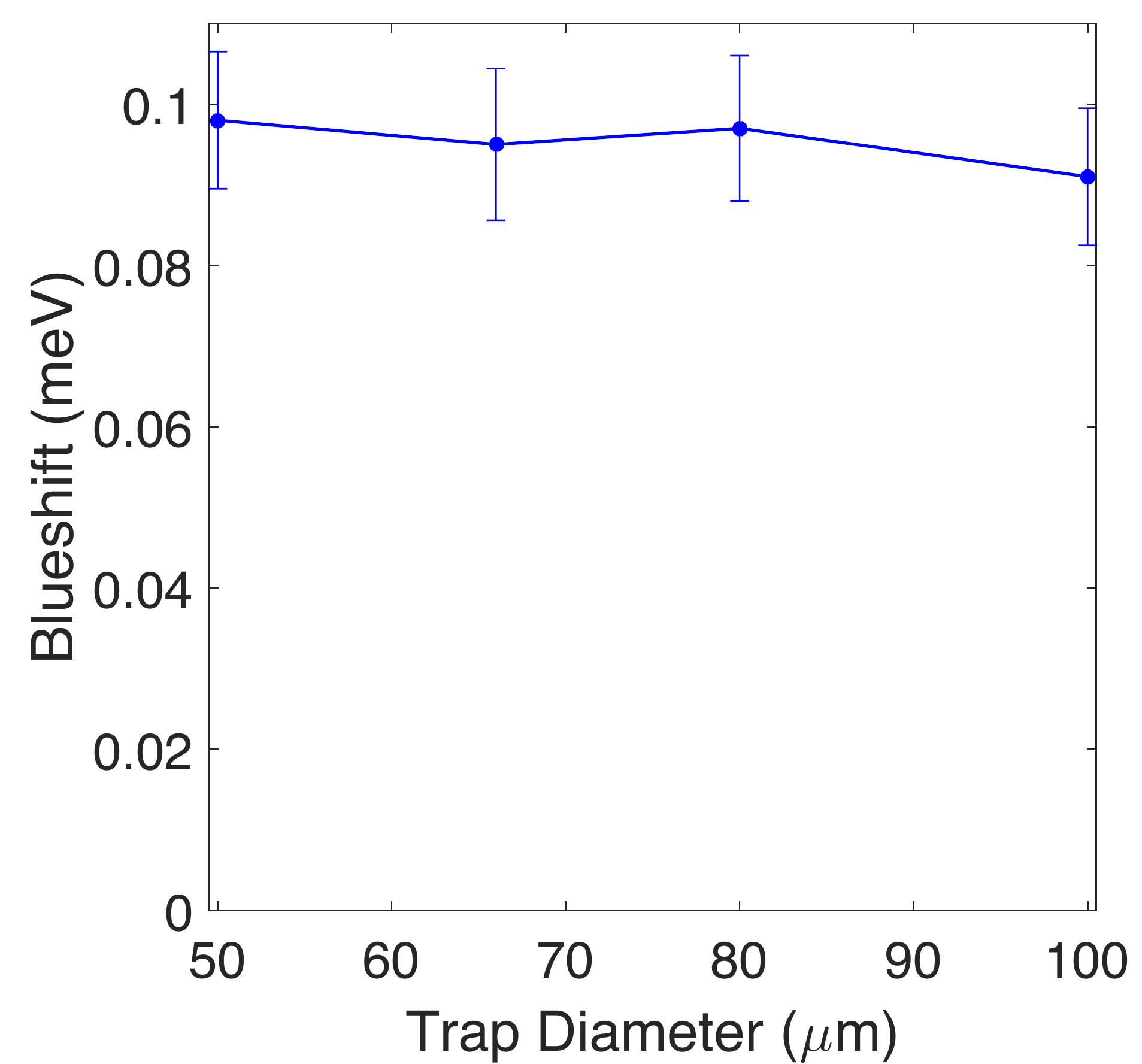}
   \caption{Blue shift of the $k_{||}=0$ state for different ring trap sizes at the cavity detuning of $\delta = -5.87$ meV when the density of the polaritons in the center of trap is $n=0.5$ $\mu$m$^{-2}$.}
\label{shift}
\end{figure}

\vspace{.5cm} 
\noindent\textbf{\textrm{Determining that excitons from the pumping region do not diffuse into the field of view}}.\hspace{10pt} It is well known that excitons can propagate up to $\sim$100 nm in typical inorganic \cite{Ramsteiner1997, Zhao2002} and organic systems \cite{Akselrod2010, Akselrod2014}. This limited range of propagation is also the main obstacle for an optical transient grating measurement of the exciton diffusion length in these systems.  Nonetheless, a careful determination whether excitons from the pumping region diffuse into the field of view has been carried out. Fig.~\ref{shift} shows the energy shifts for the ground state of the lower polaritons for different sizes of the ring trap when the density  of the polaritons in the field of view is 0.5 $\mu$m$^{-2}$ (kept constant by keeping the photon emission intensity of the polaritons constant) at a cavity detuning of $\delta = -5.87$ meV. As can be seen, when the trap size is greater than 50 $\mu$m, corresponding to a distance of 25$~\mu$m from the generation region on the perimeter to the center of the trap, the energy of the ground state remains almost the same, which confirms that the energy renormalization only comes from the contribution of polariton-polariton interactions in the center of trap, where excitons do not exist.

\begin{figure}[htbp]
\centering
  \includegraphics[width=0.60\textwidth]{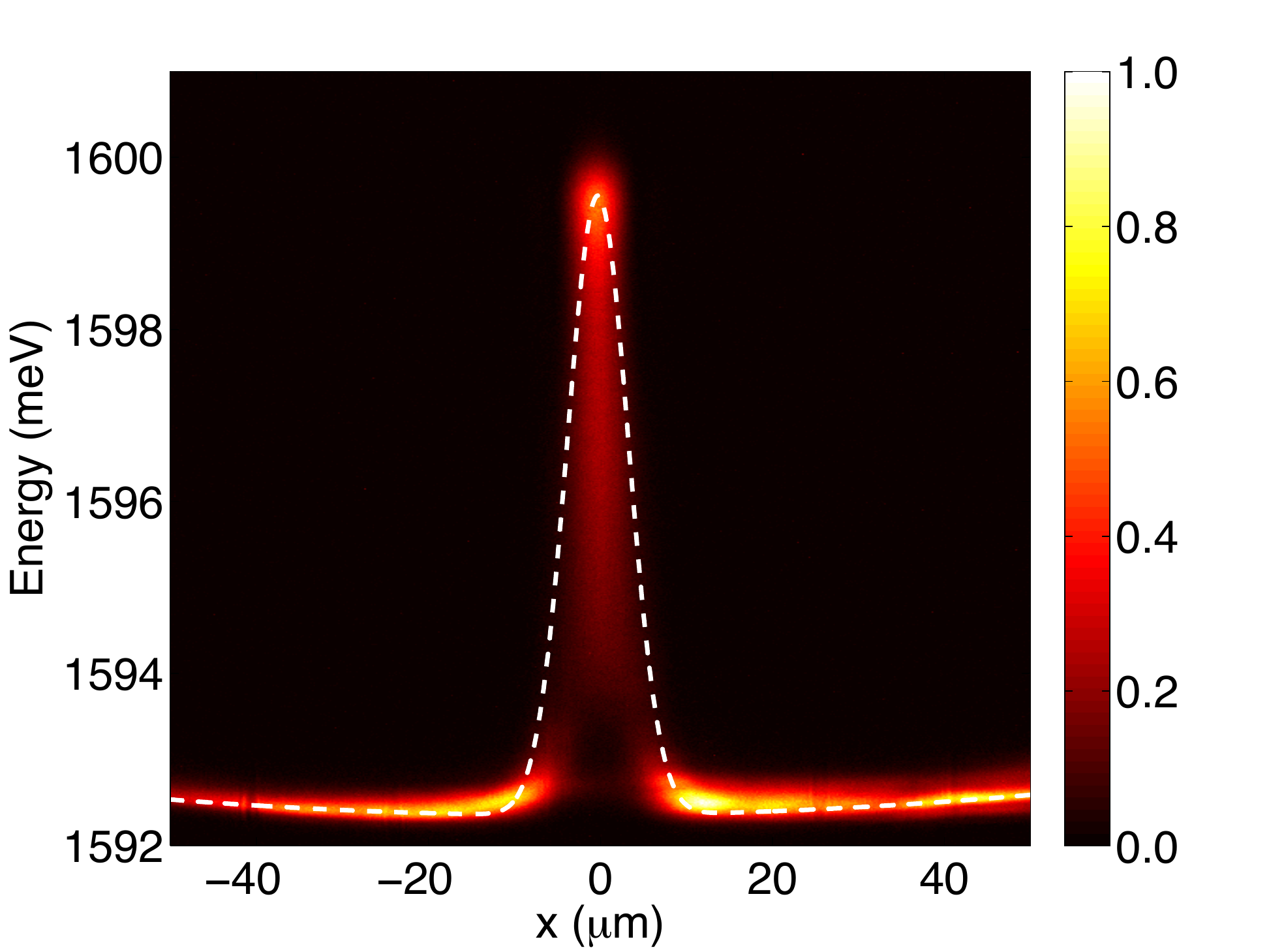}
   \caption{$k_{||}\simeq 0$ polariton energies under a Gaussian-shaped spot excitation at a cavity detuning of $\delta = 2.3$ meV in a stress trap. The white dashed line is a fit to Gaussian profile $A\exp(-x^2/\Delta x^2)$, with the diffusion length of the excitons and free carriers determined from the fit as $\Delta x= 7.5$ $\mu$m. The lattice temperature was approximately 8 K.}
\label{diffusion}
\end{figure}

The upper bound of the diffusion length of the incoherent excitons generated from non-resonant excitation has also been measured using a single excitation spot in a different trapping geometry. An excitation laser with a Gaussian intensity profile with a FWHM of approximately $8$ $\mu$m was used to excite the microcavity sample to generate a population of incoherent excitons as well as polaritons, using the same non-resonant excitation method as for the ring generation discussed in the main text, but in a stress-generated trap of the type presented in Ref.~\cite{Balili2007}.  Fig.~\ref{diffusion} shows the PL of the lower polaritons with $k_{||} = 0$. At the excitation region around $x=0$ $\mu$m, the detuning is $\delta = 2.3$~meV. The white dashed line corresponds to the single-particle energy of the lower polariton at low density in this region, found by measuring the $k_{||} = 0$ PL energy at a very low density. 
The peak at $x=0$ $\mu$m is a fit  of emission energies to a Gaussian profile $A\exp(-x^2/\Delta x^2)$. The fitting gives a FWHM of $7.5$ $\mu$m and a height of 7.2 meV. Because the generation of the polaritons is done at a single spot in this case, the excitation density must be much higher to get a number of polaritons in the trap comparable to the ring generation case.  In the case shown in Fig.~\ref{diffusion}, the polariton density is still below the condensation threshold, but the blue shift at the excitation spot is much larger than that of the excitation region the ring region for the experiments reported here; the blue shift of the excitation region was at most 4 meV in the experiments reported in the main text. The diffusion of the excitons can be seen from the energy shift of the polaritons in the vicinity of the excitation region. At $x = \pm10$ $\mu$m, the blue shift of polariton energy reduces to 0.2 meV. Based on this, the upper bound of exciton diffusion length  was estimated to be $5.3$ $\mu$m. At regions that are $> 20$ $\mu$m away from the center of the excitation region, the energy shift of the polariton ground state is negligible. 

\begin{figure}[htbp]
\centering
  \includegraphics[width=0.40\textwidth]{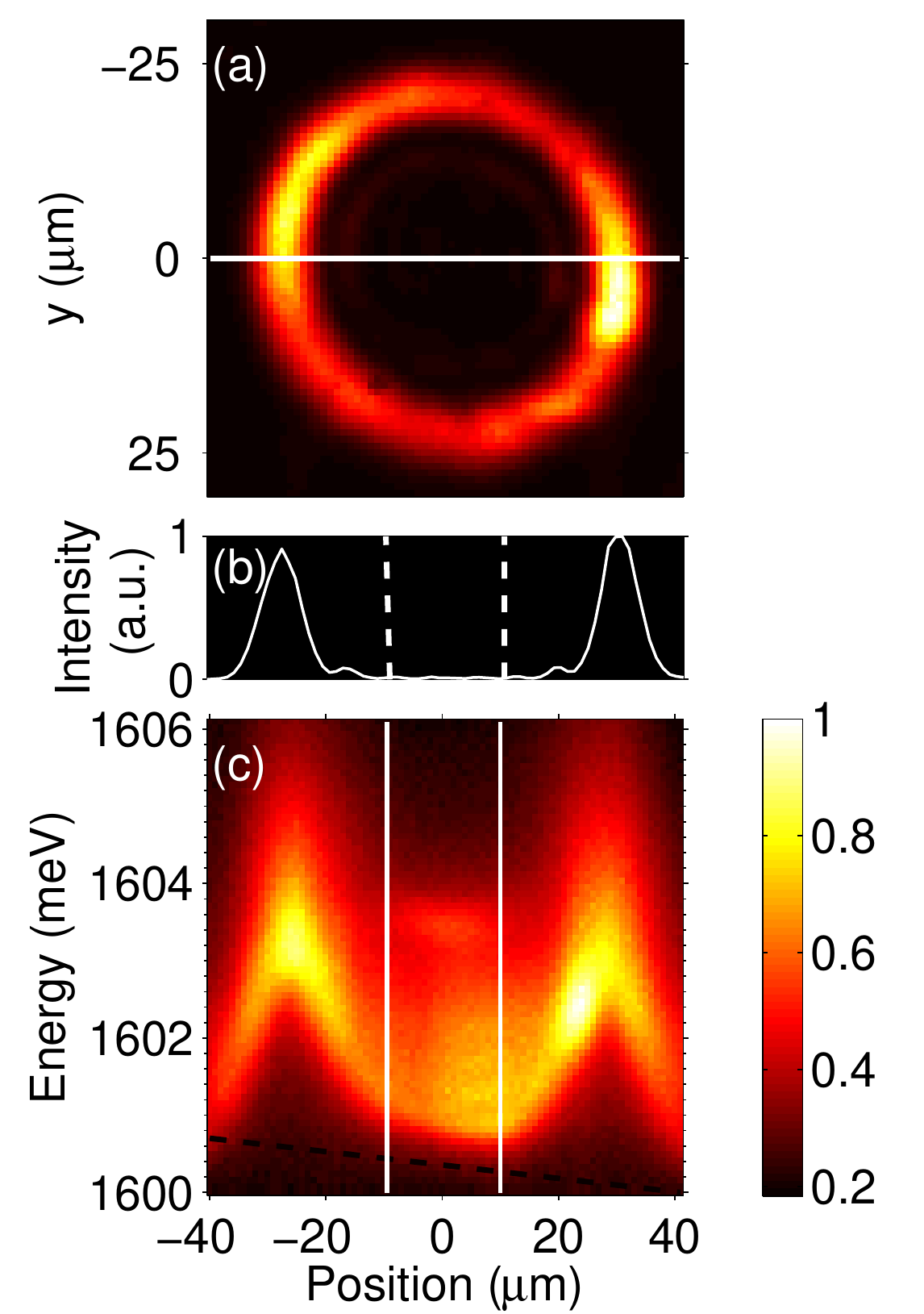}
   \caption{a) Ring profile for a different laser-generated trap with a diameter of 45 $\mu$m. b) Intensity profile for the horizontal slice shown in (a) as the white line. c) the intensity of the polariton emission as a function of energy and position for this trap over a range of angles from 0 to 15.6 degrees, corresponding to integration over the thermal $k_{||}$ states up to $k_{||}=2.2\,\mu$m$^{-1}$. The detuning at the center of the trap was $\delta = $ 2.79 meV; the bath temperature was 10 K, and the excitation laser power was $P = $ 527 mW.}
\label{smalltrap}
\end{figure}

This confirms that the region in the center of the ring trap, where the energy shifts reported in the main text were measured, was essentially free of excitons. Although this experiment with a single spot in a stress trap does not have conditions identical to those of the laser-generated ring trap, it should overestimate, not underestimate, the effect of diffusion of the excitons. First, the density at the excitation spot was higher, as evidenced by the much stronger blue shift, so the pressure pushing excitons outward should be stronger. Also, the lattice temperature was lower, which should give higher exciton diffusion constant.

Fig.~\ref{smalltrap} shows data for a ring trap with a diameter of 45 $\mu$m. In this case the excitation density is low, so there is no polariton condensate. The polaritons accumulate in the center of the trap, but do not accumulate outside the trap. The polariton density is slightly higher on the right side of the trap, as there is a net force in that direction due to the wedge of the cavity, shown as the dashed black line.

The fact that the blue shift is asymmetric, larger inside the trap than outside it, is further evidence that it does not arise from exciton diffusion. Exciton diffusion from the generation region should be symmetric, as excitons flow both inward and outward.

\begin{figure}[htbp]
\centering
  \includegraphics[width=0.50\textwidth]{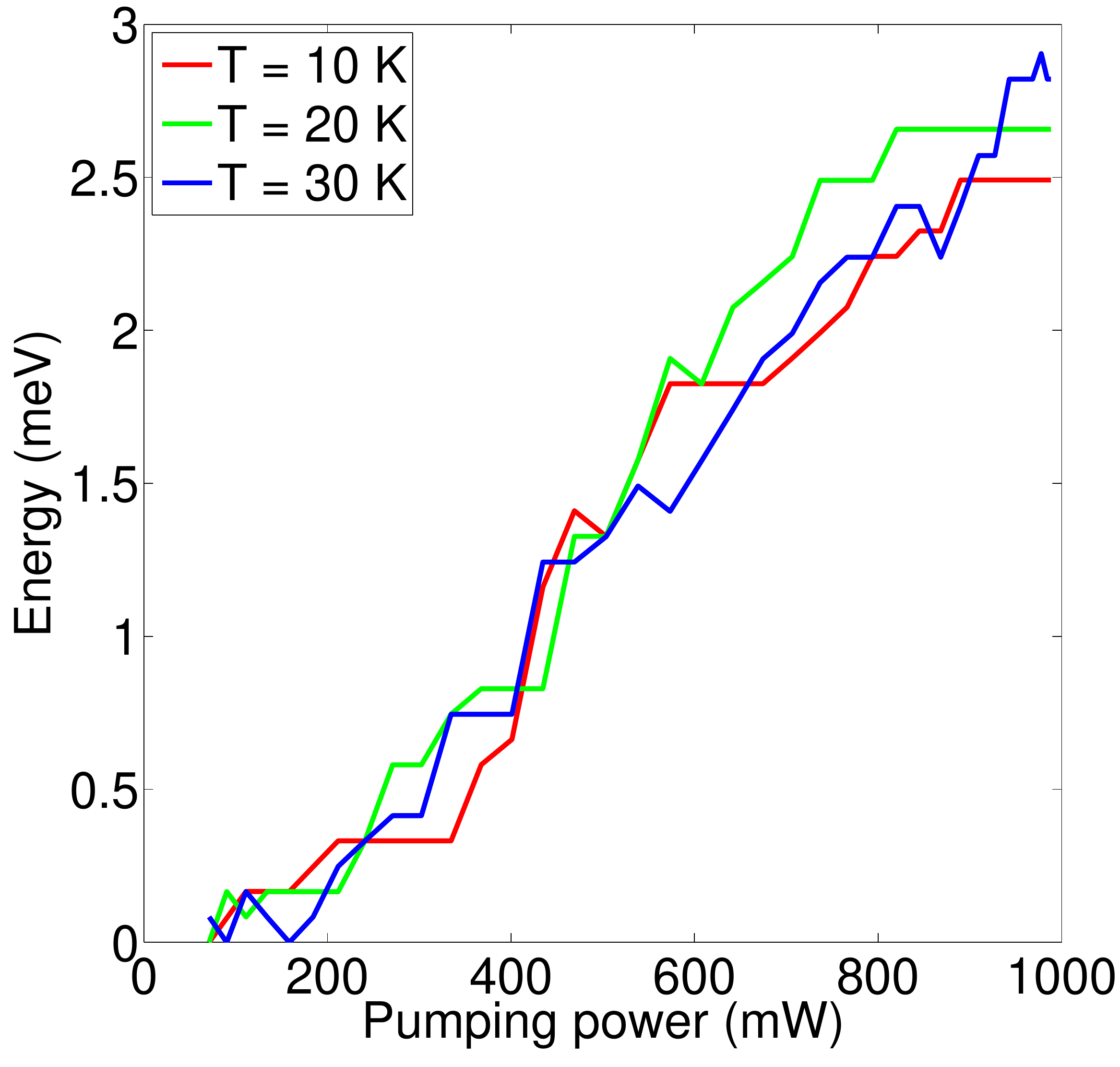}
   \caption{Blue shift of the $k_{||}=0$ state at the pumping region in an annular trap with a diameter of 66 $\mu$m as a function of pump power for different temperatures, for $\delta = -0.16 $ meV.}
\label{Tshift}
\end{figure}

\vspace{.5cm}
\noindent\textbf{Temperature dependence of the blue shift}. The temperature dependence of the blue shift under typical conditions is shown in Fig.~\ref{Tshift}. As seen in this figure, the shift at the pumping region is not strongly dependent on the lattice temperature. This is evidence against the mechanism discussed in the main text which conjectures that the polaritons generate excitons due to thermal excitation after they move to the center of the trap. Since a thermal excitation process is exponentially sensitive to the temperature due to the Boltzmann factor $e^{-\Delta E/k_BT}$, such a process would be strongly temperature dependent.

Additionally, the weak dependence on temperature also indicates that exciton diffusion from the generation region on the ring into the trap is not playing a major role, because the exciton diffusion constant should also be strongly dependent on the lattice temperature, with much higher diffusion at lower temperatures. 

On the other hand, our experiments indicate that the blue shift at the Bose-Einstein condensation threshold is roughly linear with temperature. This can be understood as arising from the density dependence of the critical temperature of the condensation threshold. In two dimensions, the critical temperature should be linear with density. The mapping of the BEC threshold density as a function of temperature will be the subject of another publication.

\end{document}